\documentclass[12pt,english,american]{article}
\usepackage[T1]{fontenc}
\usepackage[latin9]{inputenc}
\usepackage{geometry}
\usepackage{color}
\usepackage{babel}
\usepackage{amsthm}
\usepackage{amstext}
\usepackage{amssymb}
\usepackage{mathrsfs}
\usepackage{amsmath}
\usepackage{graphicx}
\usepackage{cancel}   
\usepackage[unicode=true,pdfusetitle,
 bookmarks=true,bookmarksnumbered=false,bookmarksopen=false,
 breaklinks=true,pdfborder={0 0 0},backref=false,colorlinks=true]
 {hyperref}
 \usepackage{setspace}
\hypersetup{
 linkcolor=blue,citecolor=blue, urlcolor=blue}
\makeatletter

\def\qed{$\Box$\medskip}

\newcommand{\beq}{\begin{equation}}
\newcommand{\eeq}{\end{equation}}
\newcommand{\beqa}{\begin{eqnarray}}
\newcommand{\eeqa}{\end{eqnarray}}
\newcommand{\ben}{\begin{arabicenumerate}}
\newcommand{\een}{\end{arabicenumerate}}

\def\bel{\begin{lem} } 
\def\eel{\end{lem} }
\def\bet{\begin{thm}}
\def\eet{\end{thm}}
\def\bed{\begin{defn}}
\def\eed{\end{defn} }

\def\bec{\begin{cor}}
\def\eec{\end{cor}}
\def\ber{\begin{rem}}
\def\eer{\end{rem}}

\usepackage{enumitem}		
\theoremstyle{plain}
\newtheorem{thm}{\protect\theoremname}[section]
\theoremstyle{definition}
\newtheorem{defn}[thm]{\protect\definitionname}
\theoremstyle{plain}

\theoremstyle{plain}
\theoremstyle{remark}
\newtheorem{rem}[thm]{\protect\remarkname}
\theoremstyle{plain}
\newtheorem{lem}[thm]{\protect\lemmaname}
\theoremstyle{plain}
\newtheorem{cor}[thm]{\protect\corollaryname}

\usepackage{txfonts,refstyle,xcolor}

\newref{con}{name = Conjecture\ }
\newref{prop}{name = Proposition\ }
\newref{def}{name = Definition\ }
\newref{sec}{name = Section\ }
\newref{sub}{name = Section\ }
\newref{thm}{name = Theorem\ }
\newref{lem}{name = Lemma\ }
\newref{cor}{name = Corollary\ }
\newref{fig}{name = Figure\ }
\newref{rem}{name =  Remark\ }

\usepackage{bbm}
\newcommand{\charf}{\mathbbm{1}}

\usepackage[all]{xy}

\newcommand{\xyR}[1]{%
     \makeatletter
     \xydef@\xymatrixrowsep@{#1}
     \makeatother
}

\newcommand{\xyC}[1]{%
     \makeatletter
     \xydef@\xymatrixcolsep@{#1}
     \makeatother
}

\newcommand{\ncol}[1]{\color{normalcolor}}

\makeatother

\addto\captionsamerican{\renewcommand{\corollaryname}{Corollary}}
\addto\captionsamerican{\renewcommand{\definitionname}{Definition}}
\addto\captionsamerican{\renewcommand{\lemmaname}{Lemma}}
\addto\captionsamerican{\renewcommand{\propositionname}{Proposition}}
\addto\captionsamerican{\renewcommand{\remarkname}{Remark}}
\addto\captionsamerican{\renewcommand{\theoremname}{Theorem}}
\addto\captionsenglish{\renewcommand{\corollaryname}{Corollary}}
\addto\captionsenglish{\renewcommand{\definitionname}{Definition}}
\addto\captionsenglish{\renewcommand{\lemmaname}{Lemma}}
\addto\captionsenglish{\renewcommand{\propositionname}{Proposition}}
\addto\captionsenglish{\renewcommand{\remarkname}{Remark}}
\addto\captionsenglish{\renewcommand{\theoremname}{Theorem}}
\providecommand{\corollaryname}{Corollary}
\providecommand{\definitionname}{Definition}
\providecommand{\lemmaname}{Lemma}
\providecommand{\propositionname}{Proposition}
\providecommand{\remarkname}{Remark}
\providecommand{\theoremname}{Theorem}

\addto\captionsamerican{\renewcommand{\corollaryname}{Corollary}}
\addto\captionsamerican{\renewcommand{\definitionname}{Definition}}
\addto\captionsamerican{\renewcommand{\lemmaname}{Lemma}}
\addto\captionsamerican{\renewcommand{\propositionname}{Proposition}}
\addto\captionsamerican{\renewcommand{\remarkname}{Remark}}
\addto\captionsamerican{\renewcommand{\theoremname}{Theorem}}
\addto\captionsenglish{\renewcommand{\corollaryname}{Corollary}}
\addto\captionsenglish{\renewcommand{\definitionname}{Definition}}
\addto\captionsenglish{\renewcommand{\lemmaname}{Lemma}}
\addto\captionsenglish{\renewcommand{\propositionname}{Proposition}}
\addto\captionsenglish{\renewcommand{\remarkname}{Remark}}
\addto\captionsenglish{\renewcommand{\theoremname}{Theorem}}
\providecommand{\corollaryname}{Corollary}
\providecommand{\definitionname}{Definition}
\providecommand{\lemmaname}{Lemma}
\providecommand{\propositionname}{Proposition}
\providecommand{\remarkname}{Remark}
\providecommand{\theoremname}{Theorem}

\begin{document}
\title{Lie-Schwinger block-diagonalization and gapped quantum chains with unbounded interactions} 
  \author{S. Del Vecchio\footnote{Institut f\"ur Theoretische Physik, Universit\"at Leipzig, Germany / email: simone.del\_vecchio@physik.uni-leipzig.de}\,,  J. Fr\"ohlich\footnote{Institut f\"ur Theoretiche Physik, ETH-Z\"urich , Switzerland / email: juerg@phys.ethz.ch}\,, A. Pizzo \footnote{Dipartimento di Matematica, Universit\`a di Roma ``Tor Vergata", Italy
/ email: pizzo@mat.uniroma2.it}\,, S. Rossi\footnote{Dipartimento di Matematica, Universit\`a di Roma ``Tor Vergata", Italy
/ email: stipan@alice.it}}

\date{02/12/2019}

\maketitle

\abstract{We study quantum chains whose Hamiltonians are perturbations by interactions of short range of a Hamiltonian that does not couple the degrees of freedom located at different sites of the chain and has a strictly positive energy gap above its ground-state energy. For interactions that are form-bounded  w.r.t. the on-site Hamiltonian terms, we prove that the spectral gap of the perturbed Hamiltonian above its ground-state energy  is bounded from below by a positive constant \textit{uniformly} in the length of the chain, for small values of a coupling constant. Our proof is based on an extension of a novel method introduced in \cite{FP} involving \emph{local} Lie-Schwinger conjugations of the Hamiltonians associated with connected subsets of the chain. }
\\
\section{Introduction: Models and Results}\label{intro}
In this paper, we study spectral properties of Hamiltonians of some family of quantum chains with interactions of short range, including bosonic systems, such as  an array of coupled anharmonic oscillators. We are interested in determining the multiplicity of the ground-state energy and in estimating the size of the spectral gap above the ground-state energy of Hamiltonians of such chains, as the length of the chains tends to infinity. We will consider a family of Hamiltonians for which we will prove that their ground-state energy is finitely degenerate and the spectral gap above the ground-state energy  is bounded from below by a positive constant, \textit{uniformly} in the length of the chain.
Connected sets of Hamiltonians with these properties represent what people tend to call a ``topological phase''. Recent interest in characterising topological phases of matter (see, e.g., \cite{MN}, \cite{NSY}, \cite{BN}) has motivated our analysis.

Results similar to the ones established in this paper, but mainly for bounded interactions,  have been proven before, often using so-called ``cluster expansions'': see \cite{DFF}, \cite{FFU}, \cite{KT}, \cite{Y}, \cite{KU}, \cite{DS} ,\cite{H}, \cite{MZ}, \cite{DRS} and refs. given there. Concerning bosonic systems we mention \cite{FFU} and the paper by D. Yarotsky, see \cite{Y}. In the latter paper the same type of results discussed in the following have been established in some generality for ``relatively bounded perturbations" that include the unbounded interactions discussed in Sect. \ref{intro-def}.  In particular, in \cite{Y},  small perturbations of the AKLT model have been treated by a demanding application of a cluster expansion combined with a scaling transformation.

The purpose of this paper is to extend to bosonic systems  a novel method\footnote{See \cite{DFFR} for the use of a similar block-diagonalization in a simpler context. Ideas somewhat similar to the scheme in \cite{FP}
have been used in work of J. Z. Imbrie, \cite{I1}, \cite{I2}.} described in  \cite{FP}. This method is based on \textit{iterative unitary conjugations} of the Hamiltonians, which serve to block-diagonalise them with respect to a fixed orthogonal projection 
and its orthogonal complement. We show that the algorithm employed in \cite{FP}, which  yields a flow of transformed Hamiltonians converging to a block-diagonal Hamiltonian, works for unbounded interactions, too. In this respect, the method introduced in \cite{FP} appears to be very robust, insofar as it treats fermions and bosons on the same footing. However, for the latter particles (i.e., for bosons),  the control of the algorithm requires a new proof by induction and a careful use of a weighted operator norm, combined with an iterative control of operator domains. For the problems addressed in this paper,  our technique does not face
a  \emph{large field problem},  even though bosons are involved,  in contrast to most approaches; see also Remark \ref{large-field}.

\subsection{A concrete family of quantum chains}\label{intro-def}
The Hilbert space of pure state vectors of the quantum chains studied in this paper has the form
\begin{equation}\label{tensorprod}
\mathcal{H}^{(N)}:= \bigotimes_{j=1}^{N} \mathcal{H}_{j}\,,
\end{equation}
where $\mathcal{H}_{j}\simeq \mathcal{H}, \, \forall j=1,2,\dots,$ and where $\mathcal{H}$ is a separable Hilbert space. Let $H$ be a non-negative operator on $\mathcal{H}$ with the properties that $0$ is an eigenvalue of $H$ corresponding to an eigenvector $\Omega \in \mathcal{H}$, and 
$$H \upharpoonright_{\lbrace \mathbb{C} \Omega \rbrace^{\perp}} \geq \charf \,,$$
where $\charf $  is the identity operator.
We define
\begin{equation}\label{H_i}
H_{i}:= \charf_{1}\otimes \dots \otimes \underset{\underset{i^{th} \text{slot}}{\uparrow}}{H} \otimes \dots \charf_{N}\,.
\end{equation}
By $P_{\Omega_i}$ we denote the orthogonal projector onto the subspace
\begin{equation}\label{vacuum_i}
\mathcal{H}_{1}\otimes \dots \otimes \underset{\underset{i^{th} \text{slot}}{\uparrow}}{\lbrace \mathbb{C} \Omega \rbrace}\otimes \dots \otimes \mathcal{H}_{N} \subset \mathcal{H}^{(N)}\,, \quad \text{  and}\quad   P_{\Omega_i}^{\perp} := \charf - P_{\Omega_i}\,.
\end{equation}
Then 
\begin{equation*} 
H_i = P_{\Omega_i} H_i P_{\Omega_i} + P_{\Omega_i}^{\perp} H_i P_{\Omega_i}^{\perp} \,,
\end{equation*}
with
\begin{equation}\label{gaps}
P_{\Omega_i} H_i P_{\Omega_i}=0\,,\quad P_{\Omega_i}^{\perp} H_i P_{\Omega_i}^{\perp} \geq P_{\Omega_i}^{\perp}\,.
\end{equation}
We study quantum chains on the graph $I_{N-1,1}:= \lbrace 1, \dots, N\rbrace, \,N< \infty$ arbitrary, with a Hamiltonian of the form
\begin{equation}\label{Hamiltonian}
K_{N}\equiv K_{N}(t):= \sum_{i=1}^{N} H_i + t \sum_{\underset{k \leq \bar{k}}{I_{k,i}\subset I_{N-1,1}}} V_{I_{k,i}}\,, 
\end{equation}
where $t\in \mathbb{R}$ is a coupling constant, $\bar{k} < \infty$ is an arbitrary, but fixed integer,
$I_{k,i}$ is the ``interval'' given by $\lbrace i, \dots, i+k\rbrace, \, i=1, \dots, N-k$, \, and $V_{I_{k,i}}$ is a symmetric operator acting on $\mathcal{H}^{(N)}$ with the property that 
\begin{equation}\label{potential}
V_{I_{k,i}} \,\, \text{  acts as the identity on  }\,\, \bigotimes_{j\in I_{N-1,1}\,,\, j\notin I_{k,i}} \mathcal{H}_{j}\,.
\end{equation}
The interval $I_{k,i}$ is called the ``support'' of $V_{I_{k,i}}$. Furthermore, we assume that $D((H_{I_{k,i}}^0)^{\frac{1}{2}})\subseteq D(V_{I_{k,i}})$ where $H_{I_{k,i}}^0:=\sum_{l=i}^{i+k}H_l$,  and 
for any $\phi \in D((H_{I_{k,i}}^0)^{\frac{1}{2}})$ 
\begin{equation}\label{klmn-cond}
|\langle \phi\,,\, V_{I_{k,i}} \phi \rangle|\leq a\langle \phi\,,\,(H_{I_{k,i}}^0+1)\phi\rangle,
\end{equation}
for some universal constant $a>0$. Under these assumptions,  using the inequality
\begin{equation}
\sum_{I_{k,i}\subset I_{N-1,1}}H_{I_{k,i}}^0\leq (k+1)\sum_{i=1}^{N} H_i\,,
\end{equation}
we know that for $|t|$ sufficiently small (depending on $\bar{k}$ and $a$, but independent of $N$) the symmetric operator in (\ref{Hamiltonian})  is defined and bounded from below on $D(H^0_{I_{N-1,1}})$, and can be extended to a densely defined self-adjoint operator whose domain we denote $D(K_N)\subseteq D((H_{I_{N-1,1}}^0)^{\frac{1}{2}})$, namely the Friedrichs extension of the operator in (\ref{Hamiltonian}). It is not difficult to check that,  under our hypotheses on the potentials,  this extension coincides with the self-adjoint operator defined through the KLMN theorem starting from the closed quadratic form associated with  (\ref{Hamiltonian}).

\noindent
The constraint in (\ref{klmn-cond}) readily implies 
that 
\begin{equation}\label{weighted}
\|(H_{I_{k,i}}^0+1)^{-\frac{1}{2}}V_{I_{k;i}}(H_{I_{k,i}}^0+1)^{-\frac{1}{2}}\|\leq a\,.
\end{equation}
Hence we introduce the weighted norm
\begin{equation}\label{weighted-norm}
\|V_{I_{k;i}}\|_{H^0}:=\|(H_{I_{k,i}}^0+1)^{-\frac{1}{2}}V_{I_{k;i}}(H_{I_{k,i}}^0+1)^{-\frac{1}{2}}\|\,
\end{equation}
where we point out that the weight $(H_{I_{k,i}}^0+1)^{-\frac{1}{2}}$ depends on the interval $I_{k,i}$ though this is not reflected in the symbol $\|\cdot \|_{H^0}$.
Without loss of generality, we may assume that $a=\frac{1}{2}$. 

Our results apply to anharmonic quantum crystal models described by Hamiltonians of the type
\begin{equation}
K^{crystal}_{N}:=\sum_{j=1}^{N}\Big(-\frac{d^2}{dx^2_j}+V(x_j))\Big)+t\sum_{j=1}^{N-1}W(x_j,x_{j+1})=:\sum_{j=1}^{N}H_j+t\sum_{j=1}^{N-1}W(x_j,x_{j+1})
\end{equation}
acting on the Hilbert space $\mathcal{H}^N:=\otimes_{j=1}^{N}L^2(\mathbb{R}\,,\,dx_j)$, with $V(x_j)\geq 0$,  $V(x_j)\to \infty$ for $|x_j|\to \infty$,  $D((H_j+H_{j+1})^{\frac{1}{2}})\subseteq D(W(x_j,x_{j+1}))$, and $W(x_j,x_{j+1})$ form-bounded by $H_j+H_{j+1}$.  The class described above includes the $\phi^4-$model on the one-dimensional lattice, corresponding to $V(x_j)=x_j^2+x_j^4$ and $W(x_i,x_j)=x_jx_{j+1}$.

To simplify our presentation, starting from Sect. \ref{section-gap} we consider a \emph{nearest-neighbor interaction} with 
$$\|V_{I_{1,i}}\|_{H^0}:=\|(H_{I_{1,i}}^0+1)^{-\frac{1}{2}}V_{I_{1,i}}(H_{I_{1,i}}^0+1)^{-\frac{1}{2}}\|=\frac{1}{2}\,$$ 
and $t>0$ small enough. However, with obvious modifications, our proof can be adapted to general Hamiltonians of the type in (\ref{Hamiltonian}).

\subsection{Main result}
The main result in this paper is the following theorem proven in Section \ref{norms}, (see Theorem \ref{main-res}).

{\bf{Theorem.}}
\textit{Under the assumption that (\ref{gaps}), (\ref{potential}) and (\ref{weighted}) hold, the Hamiltonian $K_{N}$ defined in (\ref{Hamiltonian}) has the following properties: There exists some $t_0 > 0$ such that, for any $t\in \mathbb{R}$ with 
$\vert t \vert < t_0$, and for all $N < \infty$,
\begin{enumerate}
\item[(i)]{ $K_{N}$ has a unique ground-state; and}
\item[(ii)]{ the energy spectrum of $K_N$ has a strictly positive gap, $\Delta_{N}(t) \geq \frac{1}{2}$, above the ground-state energy.}
\end{enumerate}
}

Results similar to the theorem stated above have appeared in the literature; see  \cite{Y}. The main novelty introduced in this paper is our method of proof.

We define
\begin{equation}\label{vacuum-proj}
P_{vac}:=\bigotimes_{i=1}^{N} P_{\Omega_i}\,.
\end{equation}
Note that $P_{vac}$ is the orthogonal projection onto the ground-state of the operator $K_{N}(t=0)=\sum_{i=1}^{N} H_{i}$. Our aim is to find an anti-symmetric operator $S_{N}(t)=-S_{N}(t)^{*}$ acting on $\mathcal{H}^{(N)}$ (so that 
exp$\big(\pm S_{N}(t)\big)$ is unitary) with the property that, after conjugation, the operator
\begin{equation}\label{conjug}
e^{S_{N}(t)}K_{N}(t)e^{-S_{N}(t)}=: \widetilde{K}_{N}(t)
\end{equation}
is \textit{``block-diagonal''} with respect to $P_{vac}$, $P_{vac}^{\perp} (:= \charf - P_{vac})$, in the sense that $P_{vac}$ projects onto the ground-state of $\widetilde{K}_{N}(t)$,
\begin{equation}\label{block-diag}
\widetilde{K}_{N}(t)= P_{vac} \widetilde{K}_{N}(t) P_{vac} + P_{vac}^{\perp} \widetilde{K}_{N}(t) P_{vac}^{\perp}\,,
\end{equation}
and 
\begin{equation}\label{gapss}
\text{infspec}\left(P_{vac}^{\perp}\widetilde{K}_{N}(t) P_{vac}^{\perp} \upharpoonright_{P_{vac}^{\perp} \mathcal{H}^{(N)}}\right)
\geq \text{infspec} \left(P_{vac} \widetilde{K}_{N}(t) P_{vac} \upharpoonright_{P_{vac}\mathcal{H}^{(N)}}\right) + \Delta_{N}(t)\,,
\end{equation}
with $\Delta_{N}(t) \geq \frac{1}{2}$, for $\vert t \vert < t_0$, \textit{uniformly} in $N$.
The iterative construction of the operator $S_{N}(t)$, yielding (\ref{block-diag}), and the proof of (\ref{gapss}) are the main tasks to be carried out. We shall block-diagonalize $K_{N}(t)$ by conjugating it iteratively by \emph{local} unitary operators chosen according to the ``Lie-Schwinger procedure''. Indeed, these operators are supported on the subsets $I_{k,q}$ of $\lbrace 1, \dots, N \rbrace$. The block-diagonalization will concern operators acting on tensor-product  spaces of the sort 
$\mathcal{H}_{q} \otimes \dots \otimes \mathcal{H}_{k+q}$ (and acting trivially on the remaining tensor factors), and it will be with respect to the projector onto the ground-state (``vacuum'') subspace, $\lbrace \mathbb{C}(\Omega_{q}\otimes \dots \otimes \Omega_{k+q})\rbrace$, contained in $\mathcal{H}_{q} \otimes \dots \otimes \mathcal{H}_{k+q}$ and its orthogonal complement. The challenge is to control the new interaction terms that are being created  along the way. They are \emph{unbounded} operators whose support corresponds to ever longer intervals (connected subsets) of the chain.

Formal aspects of our construction, along with a comparison  of bounded and unbounded interactions, respectively, are described in Sect. \ref{sect-lie} and Sect. \ref{bound-unbound}. Sect. \ref{algo} is devoted to  a detailed description of the algorithm yielding the effective potentials at each step of the block-diagonalization procedure. In Sect. \ref{norms}, the proof of convergence of our construction of the operator $S_{N}(t)$ and the proof of a lower bound, $\Delta_{N}(t)$,  on the spectral gap, for sufficiently small values of $\vert t \vert$, are presented, with a few technicalities deferred to Appendix \ref{app}.\\


{\bf{Notation}}
\\

\noindent
1) Notice that $I_{k,q}$  can also be seen as a connected one-dimensional graph with $k$ edges connecting  the $k+1$ vertices $q,1+q,\dots, k+q$, or as an ``interval'' of length $k$ whose left end-point coincides with $q$. 
\\

\noindent
2) We use the same symbol for the operator $O_j$ acting on $\mathcal{H}_j$ and the corresponding operator $$\charf_{i}\otimes\dots \otimes  \charf_{j-1}\otimes O_j \otimes \charf_{j+1}\dots \otimes \charf_l$$ acting on $ \bigotimes_{k=i}^{l} \mathcal{H}_k$, for any $i\leq j\leq l$.
\\

\noindent
3) With the symbol ``$\subset$" we denote strict inclusion, otherwise we use the symbol  ``$\subseteq$". 
\\

{\bf{Acknowledgements.}}
A.P.  thanks  the Pauli Center, Z\"urich, for hospitality in Spring 2017 when this project got started. S. D. V. and S. R. are supported  by the ERC Advanced Grant 669240 QUEST "Quantum Algebraic Structures and Models". S. D. V., A.P., and S. R. also acknowledge the MIUR Excellence Department Project awarded to the Department of Mathematics, University of Rome Tor Vergata, CUP E83C18000100006.

\setcounter{equation}{0}
\section{\emph{Local} conjugations based on Lie-Schwinger series}\label{sect-lie}

The block-diagonalization procedure for unbounded interactions treated in this paper is essentially identical to the scheme introduced in \cite{FP} for interaction potentials acting on finite dimensional Hilbert spaces. Hence the formal aspects described in the next subsection are unchanged w.r.t. \cite{FP}, but we include it for the sake of completeness and for the convenience of the reader. Yet, the rigorous control of the series yielding the intermediate Hamiltonians $K_N^{(k,q)}$ reported in (\ref{def-transf-ham})-(\ref{def-transf-ham-bis})  below and the control of the energy spectrum of the Hamiltonians $G_{I_{k;q}}$ in (\ref{def-G}) require some modifications due to the unboundedness of the potentials. In Sect. \ref{bound-unbound} and Remarks \ref{warning}, \ref{algorithm}  we will explain how to take care of these issues and why the same underlying scheme works in spite of the more singular situation treated in this paper. Given the well known complications of most methods when applied to bosonic systems, it is remarkable that our procedure works for such systems.

\subsection{Block-diagonalization: Definitions and formal aspects}\label{intro-sect}
For each $k$, we consider $(N-k)$ block-diagonalization steps, each of them associated with  a subset $I_{k,q},\, q=1, \dots, N-k$. 
The block-diagonalization of the Hamiltonian will be  with respect to the subspaces associated with the projectors in (\ref{pro-minus})-(\ref{pro-plus}), introduced below.
By $(k,q)$ we label the block-diagonalization step associated with $I_{k,q}$. We introduce an ordering amongst these steps:
\begin{equation}
(k',q') \succ (k,q)
\end{equation} 
if $k'> k$ or if $k'=k$ and $q'>q$. 

\noindent
Our original Hamiltonian is denoted by $K^{(0,N)}_N :=K_{N}(t)$. We carry out the first block-diagonalization step yielding $K_{N}^{(1,1)}$. The index $(0,N)$  is our initial choice of the index $(k,q)$: all the on-site terms in the Hamiltonian, i.e, the terms $H_i$, are block-diagonal with respect to the subspaces associated with the projectors in (\ref{pro-minus})-(\ref{pro-plus}), for $l=0$.
Our goal is to arrive at a Hamiltonian of the form
\begin{eqnarray}\label{kappa-k-q}
K_N^{(k,q)}
& :=&\sum_{i=1}^{N}H_{i}+t\sum_{i=1}^{N-1}V^{(k,q)}_{I_{1,i}}+t\sum_{i=1}^{N-2}V^{(k,q)}_{I_{2,i}}+\dots+t\sum_{i=1}^{N-k}V^{(k,q)}_{I_{k,i}} \label{def-transf-ham}\\
& &+t\sum_{i=1}^{N-k-1}V^{(k,q)}_{I_{k+1,i}}+\dots+t\sum_{i=1}^{2}V^{(k,q)}_{I_{N-2,i}}+tV^{(k,q)}_{I_{N-1,1}}\label{def-transf-ham-bis}
\end{eqnarray}
after the block-diagonalization step $(k,q)$,
with the following properties:
\begin{enumerate}
\item
For a fixed $I_{l,i}$,  the corresponding potential term changes, at each step of the block-diagonalization procedure, up to the step $(k,q)\equiv (l,i)$; hence $V^{(k,q)}_{I_{l,i}}$ is the potential term associated with the interval $I_{l,i}$ in step $(k,q)$ of the block-diagonalization, and the superscript $(k,q)$ keeps track of the changes in the potential term in step $(k,q)$. The operator $V^{(k,q)}_{I_{l,i}}$ is symmetric and acts as the identity on the spaces $\mathcal{H}_j$ for $j\neq i,i+1,\dots,i+l$; the description of how these terms are created and estimates on their (weighted) norms are deferred to Sects. \ref{algo} and \ref{norms};
\item
for all sets $I_{l,i}$ with $ (l,i)\prec (k,q)$ and for the set $I_{l,i} \equiv I_{k,q} $, the associated potential $V^{(k,q)}_{I_{l,i}}$ is block-diagonal w.r.t. the decomposition of the identity into the sum of projectors
\begin{equation}\label{pro-minus}
P^{(-)}_{I_{l,i}}:= P_{\Omega_{i}}\otimes P_{\Omega_{i+1}}\otimes \dots \otimes P_{\Omega_{i+l}}\,,
\end{equation}
\begin{equation}\label{pro-plus}
P^{(+)}_{I_{l,i}}:= (P_{\Omega_{i}}\otimes P_{\Omega_{i+1}}\otimes \dots \otimes P_{\Omega_{i+l}})^{\perp}\,.
\end{equation}
\item We warn the reader that new potentials created along the block-diagonalization process are $t$-dependent though this is not reflected in our notation.
\end{enumerate}

\begin{rem}
The term \emph{step} is used throughout the paper with two slightly different meanings: 
\begin{enumerate}
\item[i)] as \emph{level} in the block-diagonalization iteration, e.g., $K_N^{(k,q)}$ is the Hamiltonian in step $(k,q)$; 
\item[ii)] for the block-diagonalization procedure to switch from level $(k,q-1)$ to level $(k,q)$, e.g., the step $(k,q-1)\rightarrow (k,q)$.
\end{enumerate}
\end{rem}

\begin{rem}\label{remark-decomp}
It is important to notice that if $V^{(k,q)}_{I_{l,i}}$ is block-diagonal w.r.t. the decomposition of the identity into $$P^{(+)}_{I_{l,i}}+P^{(-)}_{I_{l,i}}\,,$$
i.e., $$V^{(k,q)}_{I_{l,i}}=P^{(+)}_{I_{l,i}}V^{(k,q)}_{I_{l,i}}P^{(+)}_{I_{l,i}}+P^{(-)}_{I_{l,i}}V^{(k,q)}_{I_{l,i}}P^{(-)}_{I_{l,i}}\,\,,$$  then, for $ I_{l,i} \subset I_{r,j}$,
we have that
$$P^{(+)}_{I_{r,j}}\Big[P^{(+)}_{I_{l,i}}V^{(k,q)}_{I_{l,i}}P^{(+)}_{I_{l,i}}+P^{(-)}_{I_{l,i}}V^{(k,q)}_{I_{l,i}}P^{(-)}_{I_{l,i}}\Big]P^{(-)}_{I_{r,j}}=0\,.$$
To see that the first term vanishes, we use that
\begin{equation}
P^{(+)}_{I_{l,i}}\,P^{(-)}_{I_{r,j}}=0\,,
\end{equation}
while, in the second term, we use that
\begin{equation}
P^{(-)}_{I_{l,i}}V^{(k,q)}_{I_{l,i}}P^{(-)}_{I_{l,i}}\,P^{(-)}_{I_{r,j}}=P^{(-)}_{I_{r,j}}P^{(-)}_{I_{l,i}}V^{(k,q)}_{I_{l,i}}P^{(-)}_{I_{l,i}}P^{(-)}_{I_{r,j}}
\end{equation}
and
\begin{equation}
P^{(+)}_{I_{r,j}}P^{(-)}_{I_{r,j}}=0\,.
\end{equation}

\noindent
Hence $V^{(k,q)}_{I_{l,i}}$ is also block-diagonal with respect to the decomposition of the identity into  $$P^{(+)}_{I_{r,j}}+P^{(-)}_{I_{r,j}}\,.$$
However, notice that
\begin{equation}
P^{(-)}_{I_{r,j}}\Big[P^{(+)}_{I_{l,i}}V^{(k,q)}_{I_{l,i}}P^{(+)}_{I_{l,i}}+P^{(-)}_{I_{l,i}}V^{(k,q)}_{I_{l,i}}P^{(-)}_{I_{l,i}}\Big]P^{(-)}_{I_{r,j}}
=P^{(-)}_{I_{r,j}}\,V^{(k,q)}_{I_{l,i}}\,P^{(-)}_{I_{r,j}}\,.
\end{equation}
But 
$$P^{(+)}_{I_{r,j}}\Big[P^{(+)}_{I_{l,i}}V^{(k,q)}_{I_{l,i}}P^{(+)}_{I_{l,i}}+P^{(-)}_{I_{l,i}}V^{(k,q)}_{I_{l,i}}P^{(-)}_{I_{l,i}}\Big]P^{(+)}_{I_{r,j}}$$
remains as it is.
\end{rem}
\begin{rem}\label{rem-block}
The block-diagonalization procedure that we will implement enjoys the property that the terms  block-diagonalized along the process do not change, anymore, in subsequent steps.
\\

\end{rem}
\subsection{Lie-Schwinger conjugation associated with $I_{k,q}$}\label{L-S-scheme}
Here we explain the block-diagonalization step from $(k,q-1)$ to $(k,q)$ by which the term $V^{(k,q-1)}_{I_{k,q}}$ is transformed  to a new operator, $V^{(k,q)}_{I_{k,q}}$, which is block-diagonal  w.r.t.  the decomposition of the identity into  $$P^{(+)}_{I_{k,q}}+P^{(-)}_{I_{k,q}}\,.$$
We note that the steps of the type\footnote{The initial step, $(0,N)\rightarrow (1,1)$,  is of this type; see the definitions in (\ref{initial-V}) corresponding to a Hamiltonian $K_N$ with \emph{nearest-neighbor interactions}.} $(k, N-k)\, \rightarrow \,(k+1, 1)$ are somewhat different,  because the first index (i.e., the number of edges of the interval) is changing from $k$ to $k+1$. Here we deal with general steps $(k, q-1)\, \rightarrow \,(k, q)$, with $N-k\geq q\geq 2$, and we refer the reader to \cite{FP} for the special steps mentioned above that require a slightly different notation.

\begin{rem} We warn the reader that, in the discussion below  and in Definition \ref{def-interections}, some of the steps are only formal, due to the presence of unbounded operators and of series of operators; for example the definition  in (\ref{def-AD}) might be ill posed for unbounded operators; but, as shown in the proof of Theorem \ref{th-norms},  the formula is still meaningful for the operators studied here. With regard to the identity in  (\ref{def-KN}), we remark that in Theorem \ref{th-potentials} the r-h-s will be shown to be a well defined self-adjoint operator starting from the associated quadratic form.
\end{rem}

We recall that the Hamiltonian $K_N^{(k,q-1)}$ is given by
\begin{eqnarray}
K_N^{(k,q-1)}
& :=&\sum_{i=1}^{N}H_{i}+t\sum_{i=1}^{N-1}V^{(k,q-1)}_{I_{1,i}}+t\sum_{i=1}^{N-2}V^{(k,q-1)}_{I_{2,i}}+\dots+t\sum_{i=1}^{N-k}V^{(k,q-1)}_{I_{k,i}} \\
& &+t\sum_{i=1}^{N-k-1}V^{(k,q-1)}_{I_{k+1,i}}+\dots+t\sum_{i=1}^{2}V^{(k,q-1)}_{I_{N-2,i}}+t V^{(k,q-1)}_{I_{N-1,1}}
\end{eqnarray}
and has the following properties
\begin{enumerate}
\item
each operator $V^{(k,q-1)}_{I_{l,i}}$ acts as the identity on the spaces $\mathcal{H}_j$ for $j\neq i,i+1,\dots,i+l$. In Sect. \ref{algo} we explain how these terms are created, and in Sect. \ref{norms} how their norms can be estimated;
\item
each operator $V^{(k,q-1)}_{I_{l,i}}$, with $l<k$ or $l=k$ and $q-1\geq i$,  is block-diagonal w.r.t. the decomposition of the identity into the sum of projectors in (\ref{pro-minus})-(\ref{pro-plus}).
\end{enumerate}

With the next block-diagonalization step, labeled by $(k,q)$, we want to block-diagonalize the interaction term $V^{(k,q-1)}_{I_{k,q}}$, considering the operator
\begin{equation}\label{def-G}
G_{I_{k,q}}:=\sum_{i\subset I_{k,q} }H_i+t \sum_{I_{1,i} \subset I_{k,q}} V^{(k, q-1)}_{I_{1,i}}+\dots+t \sum_{I_{k-1,i}\subset I_{k,q}}V^{(k, q-1)}_{I_{k-1,i}}\,,
\end{equation}
as the ``unperturbed" Hamiltonian. This operator is block-diagonal w.r.t. the decomposition of  the identity, i.e., 
\begin{equation}
G_{I_{k,q}}=P^{(+)}_{I_{k,q}}G_{I_{k,q}}P^{(+)}_{I_{k,q}}+P^{(-)}_{I_{k,q}}G_{I_{k,q}}P^{(-)}_{I_{k,q}}\,;
\end{equation}
see Remarks \ref{remark-decomp} and \ref{rem-block}.
We also define
\begin{equation}\label{def-E-bis}
E_{I_{k,q}}:=t\sum_{I_{1, i} \subset I_{k,q}}\langle  V^{(k,q)}_{I_{1,i}} \rangle +\dots+t\sum_{I_{k-1,i}\subset  I_{k,q}}\langle V^{(k,q)}_{I_{k-1,i}}\rangle
\end{equation}
where 
\begin{eqnarray}
\langle  V^{(k, q)}_{I_{j,i}} \rangle &:=&\langle \Omega_{i}\otimes \dots \otimes \Omega_{i+j}\,,\,V^{(k,q)}_{I_{j,i}} \,\Omega_{i}\otimes \dots \otimes \Omega_{i+j} \rangle \label{def-exp-V}\\
&=&\langle  \Omega_{i}\otimes \dots \otimes \Omega_{i+j}\,,\,(H_{I_{j,i}}^0+1)^{-\frac{1}{2}}V^{(k,q)}_{I_{j,i}} (H_{I_{j,i}}^0+1)^{-\frac{1}{2}}\, \Omega_{i}\otimes \dots \otimes \Omega_{i+j}\rangle\,,\quad \label{expectation}
\end{eqnarray}
so that $$G_{I_{k,q}}P^{(-)}_{I_{k,q}}=E_{I_{k,q}}P^{(-)}_{I_{k,q}}\,.$$
 Next, we sketch a convenient formalism used to construct our block-diagonalization operations; for further details the reader is referred to Sects. 2 and 3 of \cite{DFFR}. For operators $A$ and $B$, we define
\begin{equation}\label{def-AD}
ad\, A\,(B):=[A\,,\,B]\,,
\end{equation}
and, for $n\geq 2$,
\begin{equation}
ad^n A\,(B):=[A\,,\,ad^{n-1} A\,(B)]\,.
\end{equation}
In the block-diagonalization step $(k,q)$, we use the operator
\begin{equation}
S_{I_{k,q}}:=\sum_{j=1}^{\infty}t^j(S_{I_{k,q}})_j\,,
\end{equation}
where the terms $(S_{I_{k,q}})_j$ are defined iteratively; (notice that our definition is meaningful, since $(V^{(k,q-1)}_{I_{k,q}})_j$ depends on the operators $(V^{(k,q-1)}_{I_{k,q}})_1$ and $(S_{I_{k,q}})_r$,  with $r<j$):
\begin{itemize}
\item
\begin{eqnarray}\label{def-S-bis}
(S_{I_{k,q}})_j
&:=&\frac{1}{G_{I_{k,q}}-E_{I_{k,q}}}P^{(+)}_{I_{k,q}}\,(V^{(k, N-k)}_{I_{k,q}})_j\,P^{(-)}_{I_{k,q}}-h.c.
\end{eqnarray}
\item
$$(V^{(k,q-1)}_{I_{k,q}})_1=V^{(k,q-1)}_{I_{k,q}}\,,$$ 
and, for $j\geq 2$,\\
\vspace{0.2cm}

\begin{eqnarray}
&&(V^{(k,q-1)}_{I_{k,q}})_j\,:=\label{formula-V_j}\\
\quad& &\sum_{p\geq 2, r_1\geq 1 \dots, r_p\geq 1\, ; \, r_1+\dots+r_p=j}\frac{1}{p!}\text{ad}\,(S_{I_{k,q}})_{r_1}\Big(\text{ad}\,(S_{I_{k,q}})_{r_2}\dots (\text{ad}\,(S_{I_{k,q}})_{r_p}(G_{I_{k,q}}))\dots \Big)+\nonumber \\
&&\sum_{p\geq 1, r_1\geq 1 \dots, r_p\geq 1\, ; \, r_1+\dots+r_p=j-1}\frac{1}{p!}\text{ad}\,(S_{I_{k,q}})_{r_1}\Big(\text{ad}\,(S_{I_{k,q}})_{r_2}\dots (\text{ad}\,(S_{I_{k,q}})_{r_p}(V^{(k,q-1)}_{I_{k,q}}))\dots \Big)\, . \nonumber
\end{eqnarray}
\end{itemize}
The operator $S_{I_{k,q}}$ will turn out to be bounded and, consequently,  $e^{S_{I_{k,q}}}$ is invertible. We will prove that
\begin{equation}\label{def-KN}
K_N^{(k,q)}=e^{S_{I_{k,q}}}\,K_N^{(k, q-1)}\,e^{-S_{I_{k,q}}}\,
\end{equation}
where the l-h-s in (\ref{def-KN}) involves the effective potentials $V^{(k,q)}_{l,i}$ (see Sect. \ref{algo}) defined in such a way that,  \emph{a posteriori}, the identity above holds; see Theorem \ref{th-potentials}.
\\

\subsection{Bounded and unbounded interactions: similarities and differences in the strategy}\label{bound-unbound}

For bounded and for unbounded interactions, our strategy  to construct the Hamiltonians $K_N^{(k,q)}$  requires the following tools:
\begin{enumerate}
\item[1)] an algorithm to express each effective potential $V^{(k,q)}_{I_{l,i}}$ in terms of the potentials at the previous step, that must be consistent with the identity given in (\ref{def-KN});\
\item[2)] the control of the spectral gap of the Hamiltonian $G_{I_{k,q}}$ (above the ground state energy $E_{I_{k,q}}$) that must be strictly positive, uniformly in $(k,q)$ and in $N$;
\item[3)] a notion of ``smallness" of the operators describing the effective potentials, with the feature that the longer  the interval $I_{l,i}$ is, the smaller $V^{(k,q)}_{I_{l,i}}$ is.
\end{enumerate}
Items 1), 2),  and 3) above are related to one another,  this is the content of Section \ref{norms}. The control of the spectral gap given as an input in 3) is studied in Sect. \ref{section-gap}. It can be considered the core idea of our method. The algorithm we allude to in 1) is essentially the same for bounded and for unbounded interactions. But it is used in a different way in the main inductive proof (Theorem \ref{th-norms}).  We postpone a more detailed comment on this aspect to Remark \ref{algorithm}.  

In the remaining part of this section we try to explain in words the contents of  Lemma \ref{unboundedlemmaA3}, that is how the ``smallness" mentioned in 3) must be used to control the formal sums defining the operators $(V^{(k,q-1)}_{I_{k,q}})_j$ and $(S_{I_{k,q}})_j$, and the series defining $V^{(k,q)}_{I_{k,q}}$ and $S_{I_{k,q}}$. 
For bounded interactions, it is enough to show by induction that a bound of the type
\begin{equation}
\|V^{(k,q-1)}_{I_{k,q}}\|\leq \mathcal{O}(|t|^{\frac{k-1}{4}})
\end{equation}
suffices to derive an analogous bound for $\|V^{(k,q)}_{I_{k,q}}\|$, i.e.,  $\|V^{(k,q)}_{I_{k,q}}\|\leq \mathcal{O}(|t|^{\frac{k-1}{4}})$. In the present paper,  the potentials  (\ref{potential}) appearing in the original Hamiltonian $K_N$ (see (\ref{Hamiltonian})) are unbounded operators. However, we assume that they are bounded in the norm $\|\cdot \|_{H^0}$.  We can expect that the effective potentials are unbounded too,  but,  in order to have a consistent block-diagonalization scheme,  we need to prove that they are relatively bounded similarly to the potentials in the original Hamiltonian. As proven in  Lemma \ref{unboundedlemmaA3}, an assumption of the type
\begin{equation}\label{norm-bound}
\|V^{(k,q-1)}_{I_{k,q}}\|_{H^0}\leq \mathcal{O}(|t|^{\frac{k-1}{4}})
\end{equation}
implies that
\begin{enumerate}
\item[1)] the operators $(V^{(k,q-1)}_{I_{k,q}})_j$ and $V^{(k,q)}_{I_{k;q}}$ are bounded in the norm $\|\cdot \|_{H^0}$;
\item[2)] the operator $S_{I_{k;q}}$  is a bounded operator, uniformly in $k$ and $q$.  
\end{enumerate}
The more regular behaviour of $S_{I_{k;q}}$ is due to the projectors entering the definition of $(S_{I_{k,q}})_j$, since one of them, $P^{(-)}_{I_{k,q}}$, is of finite rank. 

We stress that, in  the next sections, a norm bound of the type (\ref{norm-bound}), (see  (\ref{ass-2}) below), plays a crucial role to prove that
\begin{enumerate}
\item the effective potentials $V^{(k,q)}_{I_{l,i}}$ are symmetric operators;
\item the Hamiltonian $G_{I_{k,q}}$ has a spectral gap above its ground-state energy;
\item the definitions associated with the algorithm $\alpha_{I_{k,q}}$ hold in the sense of quadratic forms.
\end{enumerate}
The rationale of the entire proof is to derive the bound in (\ref{ass-2}) from the control of the gap and from the other consequences mentioned above.
\begin{rem}\label{large-field}
Formulae (\ref{def-S-bis}) and (\ref{formula-V_j}) indicate why a \emph{large field problem} does not arise. The finite rank projector, $P^{(-)}_{I_{k,q}}$, in (\ref{def-S-bis}) has the effect to make  $S_{I_{k;q}}$ bounded (as proven in Lemma \ref{unboundedlemmaA3}), and this feature, through formula (\ref{formula-V_j}),  yields for $V^{(k,q)}_{I_{k;q}}$ the same \emph{unboundedness} as for $V^{(k,q-1)}_{I_{k;q}}$, by which we mean that the new potential (i.e., $V^{(k,q)}_{I_{k;q}}$) is bounded in the weighted operator norm, provided the preceding one (i.e., $V^{(k,q-1)}_{I_{k;q}}$) enjoys this property; see Lemma \ref{unboundedlemmaA3}.
\end{rem}

\subsubsection{Gap of the local Hamiltonians $G_{I_{k,q}}$: Main argument}\label{section-gap}
From now on, in order to simplify our presentation, we consider a \emph{nearest-neighbor interaction} with 
$$\|V_{I_{1;i}}\|_{H^0}:=\|(H_{I_{1,i}}^0+1)^{-\frac{1}{2}}V_{I_{1,i}}(H_{I_{1,i}}^0+1)^{-\frac{1}{2}}\|=\frac{1}{2}\,,$$ 
and we choose $t>0$ small enough. (However, with obvious modifications, our proof can be adapted to general Hamiltonians of the type in (\ref{Hamiltonian}).)
\\

\noindent
Our inductive hypothesis is that
\begin{equation}\label{ass-2}
\|(H_{I_{l,i}}^0+1)^{-\frac{1}{2}}V^{(k,q-1)}_{I_{l,i}}(H_{I_{l,i}}^0+1)^{-\frac{1}{2}}\|=:\|V^{(k,q-1)}_{I_{l,i}}\|_{H^0} \leq t^{\frac{l-1}{4}}\,.
\end{equation} 
The key mechanism underlying our method,  starting from the potential terms $V^{(k,q-1)}_{I_{1,i}}$, is to establish (\ref{ass-2}) by induction; see Theorem \ref{th-norms}. According to the scheme described in Section \ref{L-S-scheme},  for any  $k>1$,  the operator $V^{(k,q-1)}_{I_{1,i}}$ is block-diagonalized, i.e., 
\begin{equation}\label{informal-in}
V^{(k,q-1)}_{I_{1,i}}=P^{(+)}_{I_{1,i}}V^{(k,q-1)}_{I_{1,i}}P^{(+)}_{I_{1,i}}+P^{(-)}_{I_{1,i}}V^{(k,q-1)}_{I_{1,i}}P^{(-)}_{I_{1,i}}\,.
\end{equation}
Hence  we can  write
\begin{eqnarray}
& &P^{(+)}_{I_{k,q}}\,\Big[\sum_{i\subset I_{k,q} }H_i+t\sum_{I_{1,i} \subset I_{k,q}} V^{(k,q-1)}_{I_{1,i}}\Big]P^{(+)}_{I_{k,q}}\\
&=&P^{(+)}_{I_{k,q}}\,\Big[\sum_{i\subset I_{k,q} }H_i+t\sum_{I_{1,i} \subset I_{k,q}} P^{(+)}_{I_{1,i}}V^{(k,q-1)}_{I_{1,i}}P^{(+)}_{I_{1,i}}+t\sum_{I_{1,i} \subset I_{k,q}} P^{(-)}_{I_{1,i}}V^{(k,q-1)}_{I_{1,i}}P^{(-)}_{I_{1,i}}\Big]P^{(+)}_{I_{k,q}}\,.\quad\quad\quad \label{1.57}
\end{eqnarray}
In general, for $\psi \in D((H_{I_{r,i}}^0)^{\frac{1}{2}})$ we estimate
\begin{eqnarray}
& &|\langle \psi\,,\,P^{(+)}_{I_{r,i}}V^{(k,q-1)}_{I_{r,i}}P^{(+)}_{I_{r,i}} \psi \rangle| \label{in-est-V}\\
&= &|\langle \psi\,,\,P^{(+)}_{I_{r,i}}\,(H_{I_{r,i}}^0)^{\frac{1}{2}}(\frac{H_{I_{r,i}}^0+1}{H_{I_{r,i}}^0})^{\frac{1}{2}}(H_{I_{r,i}}^0+1)^{-\frac{1}{2}}V^{(k,q-1)}_{I_{r,i}}(H_{I_{r,i}}^0+1)^{-\frac{1}{2}}(\frac{H_{I_{r,i}}^0+1}{H_{I_{r,i}}^0})^{\frac{1}{2}}(H_{I_{r,i}}^0)^{\frac{1}{2}}\,P^{(+)}_{I_{r;i}}\psi \rangle| \quad\quad \quad \\
&\leq  &2\cdot t^{\frac{r-1}{4}}\,\langle \psi\,,\,P^{(+)}_{I_{r,i}}\,H_{I_{r,i}}^0\,P^{(+)}_{I_{r,i}}\psi \rangle\\
&\leq &2\cdot t^{\frac{r-1}{4}}\,\langle \psi\,,\,H_{I_{r,i}}^0\psi \rangle \label{fin-est-V}
\end{eqnarray}
where we have used the assumption in (\ref{ass-2}) and
\begin{equation}
\|P^{(+)}_{I_{r,i}}\,(\frac{H_{I_{r,i}}^0+1}{H_{I_{r,i}}^0})^{\frac{1}{2}}\|\leq \sqrt{2},
\end{equation}
which follows from (\ref{gaps}).

\noindent
Next,  for $1\leq l \leq L \leq N-r$, we observe that
\begin{equation}\label{ineq-inter-00}
\sum_{i=l}^{L}H^{0}_{I_{r,i}}\leq (r+1) \sum_{i=l}^{L+r} H_i \,
\end{equation}
and, using the inequality proven  in Corollary \ref{op-ineq-2}, combined with (\ref{gaps}), we find that
\begin{equation}\label{ineq-inter-0}
\sum_{i=l}^{L}P^{(+)}_{I_{r,i}}\leq (r+1) \sum_{i=l}^{L+r} P^{\perp}_{\Omega_i}\leq (r+1) \sum_{i=l}^{L+r} H_i\,.
\end{equation}
Due to the estimate in (\ref{in-est-V})-(\ref{fin-est-V}),  and using inequality (\ref{ineq-inter-00}) with $r=1$, $l=q$, $L=k+q-r$,  we have that
\begin{equation}\label{ineq-cor}
\pm \sum_{I_{1,i} \subset I_{k,q}} P^{(+)}_{I_{1,i}}V^{(k,q-1)}_{I_{1,i}}P^{(+)}_{I_{1,i}}\leq 4 \,\sum_{i=q}^{k+q} H_i\,.
\end{equation}
Hence, recalling that $t>0$ and combining (\ref{ass-2}) with (\ref{ineq-cor}), we conclude that
\begin{eqnarray}
(\ref{1.57})
&\geq & P^{(+)}_{I_{k,q}}\,\Big[(1- 4t)\,\sum_{i=q}^{k+q} H_i\Big]P^{(+)}_{I_{k,q}}+ P^{(+)}_{I_{k,q}}\,\Big[t\sum_{I_{1,i} \subset I_{k,q}}P^{(-)}_{I_{1,i}} V^{(k,q-1)}_{I_{1,i}}P^{(-)}_{I_{1,i}}\Big]P^{(+)}_{I_{k,q}}\label{second-line}\\
&=& P^{(+)}_{I_{k,q}}\,\Big[(1-4t)\sum_{i=q}^{k+q} H_i\Big]P^{(+)}_{I_{k,q}}+ P^{(+)}_{I_{k,q}}\,\Big[t\sum_{I_{1,i} \subset I_{k,q}}\langle  V^{(k,q-1)}_{I_{1,i}} \rangle P^{(-)}_{I_{1,i}}\Big]P^{(+)}_{I_{k,q}}\,,\label{third-line}
\end{eqnarray}
where $ \langle  V^{(k, q-1)}_{I_{1,i}} \rangle $ is defined in  (\ref{def-exp-V}).

Next, substituting $P^{(-)}_{I_{1,i}}=\charf -P^{(+)}_{I_{1,i}}$ into (\ref{third-line}), we get
\begin{eqnarray}
(\ref{1.57})
&\geq & P^{(+)}_{I_{k,q}}\,\Big[(1-4t)\sum_{i=q}^{k+q}H_i-t\sum_{I_{1,i} \subset I_{k,q}}\langle  V^{(k,q-1)}_{I_{1,i}} \rangle P^{(+)}_{I_{1,i}}\Big]P^{(+)}_{I_{k,q}} \quad\quad\quad\quad  \label{fin-1}\\
& &+ P^{(+)}_{I_{k,q}}\,\Big[t\sum_{I_{1,i} \subset I_{k,q}}\langle  V^{(k,q-1)}_{I_{1,i}} \rangle \Big]P^{(+)}_{I_{k,q}}\\
&\geq&P^{(+)}_{I_{k,q}}\,\Big[(1-8t)\sum_{i=q}^{k+q}H_i\Big]P^{(+)}_{I_{k,q}}\label{fin-2}\\
& &+ P^{(+)}_{I_{k,q}}\,\Big[t\sum_{I_{1,i} \subset I_{k,q}}\langle  V^{(k,q-1)}_{I_{1,i}} \rangle \Big]P^{(+)}_{I_{k,q}}\,,\label{informal-fin}
\end{eqnarray}
where, in the step\footnote{The expression $-t\sum_{I_{1,i} \subset I_{k,q}}\langle  V^{(k,q-1)}_{I_{1,i}} \rangle P^{(+)}_{I_{1,i}}$ could be actually bounded (from below) by $-2t\sum_{i=q}^{k+q}H_i$.} from (\ref{fin-1}) to (\ref{fin-2}), we have used (\ref{ineq-inter-0}), (\ref{expectation}), and (\ref{ass-2}).
Iteration of this argument yields the following lemma.
\begin{lem}\label{gap}
Assuming the bound in (\ref{ass-2}), and choosing $t>0$ so small that
\begin{equation}
1-8t-4t \sum_{l=3}^{\infty}l\cdot t^{\frac{l-2}{4}}>0\,,
\end{equation}
the inequality
\begin{equation}
P^{(+)}_{I_{k,q}}(G_{I_{k,q}}-E_{I_{k,q}})P^{(+)}_{I_{k,q}}
\geq\Big(1-8t-4t \sum_{l=3}^{\infty}l\cdot t^{\frac{l-2}{4}}\Big)H^{0}_{I_{k,q}}\,\,P^{(+)}_{I_{k,q}} \label{final-eq-1}
\end{equation}
holds in the form sense in the domain $D((H_{I_{k,q}}^0)^{\frac{1}{2}})$, where $E_{I_{k,q}}$ is defined in (\ref{def-E-bis}).
\end{lem}

\noindent
\emph{Proof}
Proceeding as in (\ref{informal-in})-(\ref{informal-fin}), hence using inequalities analogous to (\ref{ineq-cor}), i.e., \begin{equation}\label{ineq-cor-bis}
\pm \sum_{I_{l;i} \subset I_{k,q}} P^{(+)}_{I_{l,i}}V^{(k,q-1)}_{I_{l,i}}P^{(+)}_{I_{l,i}}\leq (l+1)\cdot \,\sum_{i=q}^{k+q} H_i\,,
\end{equation} we get that
\begin{eqnarray}
& &P^{(+)}_{I_{k;q}}G_{I_{k,q}}P^{(+)}_{I_{k,q}} \\
&\geq &P^{(+)}_{I_{k,q}}\,\Big[ \Big(1-4t-2t \sum_{l=3}^{k} l\cdot t^{\frac{l-2}{3}} \Big)\,\sum_{i=q}^{k+q}H_i\Big]P^{(+)}_{I_{k,q}} \quad\quad\quad\quad \nonumber \\
& &+ P^{(+)}_{I_{k,q}}\,\Big[t\sum_{I_{1;i} \subset I_{k,q}}\langle  V^{(k,q-1)}_{I_{1,i}} \rangle  P^{(-)}_{I_{1,i}}+\dots+t\sum_{I_{k-1;i}\subset  I_{k,q}}\langle V^{(k,q-1)}_{I_{k-1,i}}\rangle P^{(-)}_{I_{k-1,i}} \Big]P^{(+)}_{I_{k,q}} \,.\quad\quad\quad 
\end{eqnarray}
Next, using the identities $P^{(-)}_{I_{j,i}}+P^{(+)}_{I_{j,i}}=\charf $, 
\begin{eqnarray}
& &P^{(+)}_{I_{k,q}}G_{I_{k,q}}P^{(+)}_{I_{k,q}} \\
&\geq &P^{(+)}_{I_{k,q}}\,\Big[  \Big(1-4t-2t \sum_{l=3}^{k}l\cdot t^{\frac{l-2}{4}} \Big)\,\,\sum_{i=q}^{k+q} H_i\Big]P^{(+)}_{I_{k,q}} \nonumber \\
& &+ P^{(+)}_{I_{k,q}}\,\Big[-t\sum_{I_{1;i} \subset I_{k,q}}\langle  V^{(k,q-1)}_{I_{1,i}} \rangle  P^{(+)}_{I_{1,i}}+\dots-t\sum_{I_{k-1;i}\subset  I_{k,q}}\langle V^{(k,q-1)}_{I_{k-1,i}}\rangle P^{(+)}_{I_{k-1,i}} \Big]P^{(+)}_{I_{k,q}} \nonumber \\
& &+ P^{(+)}_{I_{k,q}}\,\Big[t\sum_{I_{1;i} \subset I_{k,q}}\langle  V^{(k,q-1)}_{I_{1,i}} \rangle +\dots+t\sum_{I_{k-1,i} \subset  I_{k;q}}\langle V^{(k,q-1)}_{I_{k-1,i}}\rangle \Big]P^{(+)}_{I_{k,q}}\,. \quad\quad\quad 
\quad\quad\quad\quad\quad\quad 
\end{eqnarray}
Finally, by using (\ref{ineq-inter-0}), we conclude that
\begin{eqnarray}
& &P^{(+)}_{I_{k,q}}G_{I_{k,q}}P^{(+)}_{I_{k,q}} \\
&\geq &P^{(+)}_{I_{k,q}}\,\Big[  \Big(1-8t-4t \sum_{l=3}^{k}l\cdot t^{\frac{l-2}{4}} \Big)\,\sum_{i=q}^{k+q} H_i\Big]P^{(+)}_{I_{k,q}} \nonumber \\
& &+ P^{(+)}_{I_{k,q}}\,\Big[t\sum_{I_{1,i} \subset I_{k,q}}\langle  V^{(k,q-1)}_{I_{1,i}} \rangle +\dots+t\sum_{I_{k-1,i} \subset  I_{k,q}}\langle V^{(k,q-1)}_{I_{k-1,i}}\rangle \Big]P^{(+)}_{I_{k,q}} \quad\quad\quad \\
&= & \Big(1-8t-4t \sum_{l=3}^{k}l\cdot t^{\frac{l-2}{4}} \Big)H^{0}_{I_{k,q}}\,\,P^{(+)}_{I_{k,q}}+ P^{(+)}_{I_{k,q}}\,\Big[t\sum_{I_{1,i} \subset I_{k,q}}\langle  V^{(k,q-1)}_{I_{1,i}} \rangle +\dots+t\sum_{I_{k-1,i} \subset  I_{k,q}}\langle V^{(k,q-1)}_{I_{k-1,i}}\rangle \Big]P^{(+)}_{I_{k,q}}\nonumber \,\\
&=&\Big(1-8t-4t \sum_{l=3}^{k}l\cdot t^{\frac{l-2}{4}} \Big)H^{0}_{I_{k,q}}\,\,P^{(+)}_{I_{k;q}}+ E_{I_{k;q}}P^{(+)}_{I_{k,q}},\nonumber 
\end{eqnarray}
where, in the last step,  we have used the definition in (\ref{def-E-bis}).
\qed

\noindent
\begin{rem}(\emph{Self-adjointness of $G_{I_{k,q}}$})\label{self}

We observe that under  assumption (\ref{ass-2}), for $t>0$ sufficiently small, but independent of $N$, $k$,  and $q$, we can extend the symmetric\footnote{In Sect. \ref{intro-sect} we have claimed that  the effective potentials $V^{(k, q-1)}_{I_{j,i}}$ are symmetric. This can be proven starting   from the algorithm $\alpha_{I_{k,q}}$  (described in  Sect.  \ref{algo})  with the assumption in (\ref{ass-2}), as explained in Remark \ref{symmetric} where we deduce that the effective potentials $V^{(k, q-1)}_{I_{j,i}}$, $I_{j,i}\subset I_{k,q}$, are symmetric in the domain $D((H^0_{I_{k,q}})^{\frac{1}{2}})$.} operator $G_{I_{k,q}}$ initially defined on $D(H^0_{I_{k,q}})$ to a self-adjoint operator, using the same argument as in Sect. \ref{intro-def} for the operator $K_N$. We keep the same notation for the self-adjoint extension, and we refer to it as the Hamiltonian  $G_{I_{k,q}}$.
\end{rem}

Hence Lemma \ref{gap} implies that, under  assumption (\ref{ass-2}),  the Hamiltonian $G_{I_{k,q}}$ has a spectral gap above its ground-state energy that can be estimated from below by $\frac{1}{2}$, for $t$ sufficiently small but \textit{independent} of $N$,  $k$, and $q$, as  stated in the Corollary below.
\begin{cor}\label{cor-gap}
Under  assumption (\ref{ass-2}), for $t>0$ sufficiently small but independent of $N$, $k$,  and $q$,  the Hamiltonian $G_{I_{k,q}}$  has  a spectral gap above the ground-state energy (given  by $E_{I_{k,q}}$ and defined in (\ref{def-E-bis})) that we estimate from below by $\Delta_{I_{k,q}}\geq \frac{1}{2}$.  The ground-state of $G_{I_{k;q}}$ coincides with the ``vacuum'', $\bigotimes_{j\in I_{k,q}}\Omega_{j}$\,\,, in $\mathcal{H}_{I_{k,q}}$,  and we have the identity
\begin{eqnarray}
P^{(-)}_{I_{k,q}}G_{I_{k,q}}P^{(-)}_{I_{k,q}}
&= &P^{(-)}_{I_{k,q}}\, \Big[t\sum_{I_{1,i} \subset I_{k,q}}\langle  V^{(k,q-1)}_{I_{1,i}} \rangle P^{(-)}_{I_{1,i}} +\dots+t\sum_{I_{k-1;i}\subset  I_{k;q}}\langle V^{(k,q-1)}_{I_{k-1,i}}\rangle P^{(-)}_{I_{k-1,i}}  \Big]P^{(-)}_{I_{k,q}}\nonumber\\
&= &P^{(-)}_{I_{k,q}}\, \Big[t\sum_{ I_{1,i}\subset I_{k,q}}\langle  V^{(k,q-1)}_{I_{1,i}} \rangle+\dots+t\sum_{I_{k-1,i}\subset  I_{k;q}}\langle V^{(k,q-1)}_{I_{k-1,i}}\rangle \Big]P^{(-)}_{I_{k,q}}\,.\label{final-eq-2}
\end{eqnarray}
\end{cor}

\section{The algorithm $\alpha_{I_{k,q}}$}\label{algo}

Here we address the question of how the interaction terms evolve under our block-diagonaliza-\\tion steps. Similarly to \cite{FP}, we define and control an algorithm, $\alpha_{I_{k,q}}$, determining a map that sends each operator $V^{(k,q-1)}_{I_{l,i}}$ to a corresponding potential term supported on the same interval, but at the next block-diagonalization step, i.e., 

\begin{equation}
\alpha_{I_{k,q}}(V^{(k,q-1)}_{I_{l,i}})=:V^{(k,q)}_{I_{l,i}}\,, 
\end{equation}
 in terms of the operators, $V^{(k,q-1)}_{I_{l,i}}$, at the previous step $(k,q-1)$, starting from 
\begin{equation}\label{initial-V}
V_{I_{0,i}}^{(0,N)}\equiv H_i\quad ,\quad
V_{I_{1,i}}^{(0,N)}\equiv V_{I_{1,i}}\quad,\quad
V_{I_{l,i}}^{(0,N)} =0\,\,\text{for}\,\, l\geq 2,
\end{equation}
and such that the identity $K_N^{(k,q)}=e^{S_{I_{k,q}}}\,K_N^{(k,q-1)}\,e^{-S_{I_{k,q}}}$ holds, where  (see (\ref{kappa-k-q})) $K_N^{(k,q)}$ and $K_N^{(k,q-1)}$ are functions of the potentials $V^{(k,q)}_{I_{l,i}}$ and $V^{(k,q-1)}_{I_{l,i}}$, respectively.

\begin{rem}\label{warning}
Given that the interaction potentials are unbounded operators, some of the steps in Definition \ref{def-interections}  are only formal. In Remark \ref{symmetric} it will be shown that the definitions hold in terms of quadratic forms.
\end{rem}

\begin{figure}
 \includegraphics[width=\linewidth]{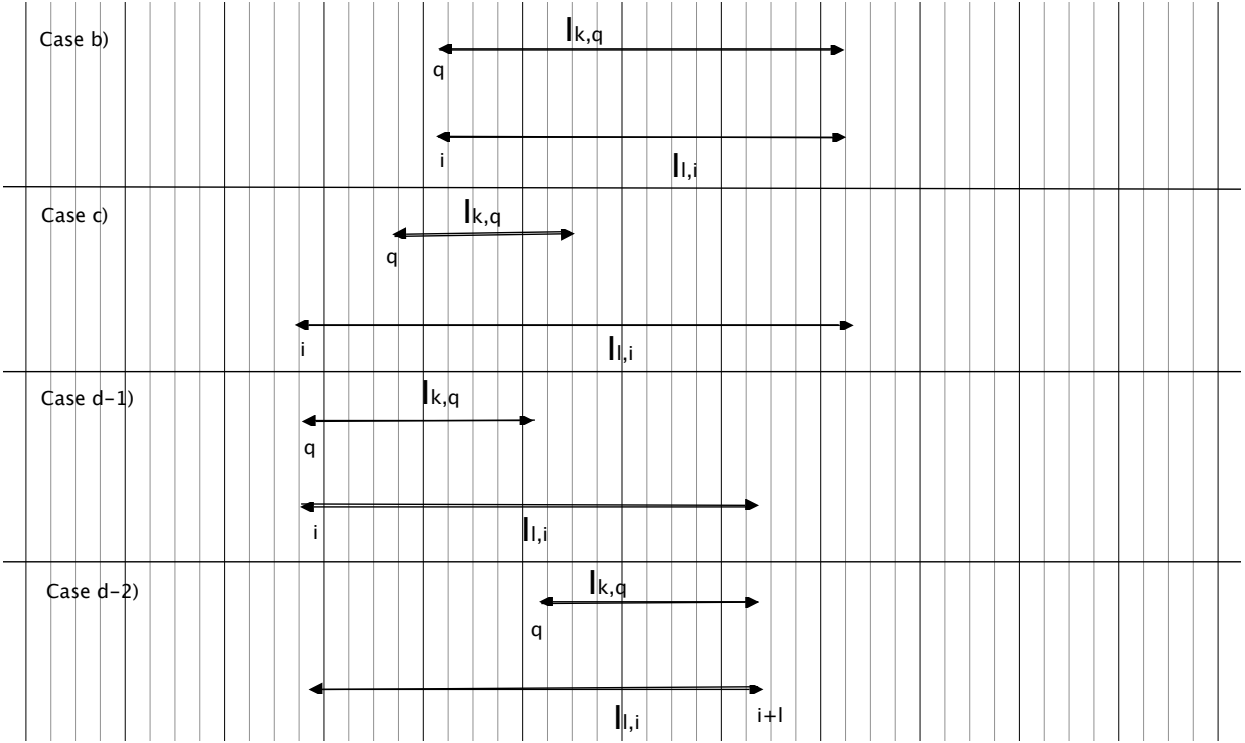}
 \caption{Relative positions of intervals $I_{k,q}$ and $I_{l,i}$}
 \label{fig:cases-bis}
\end{figure}

\begin{defn}\label{def-interections}
We assume that, for fixed $(k,q-1)$, with $(k,q-1) \succ (0,N)$, the operators $V^{(k,q-1)}_{I_{l,i}}$ and $S_{I_{k,q}}$ are well defined, for any $l,i$; or  we assume that $(k,q)=(1,1)$ and that the operator $S_{I_{1,1}}$ is well defined. We then define the operators $V^{(k,q)}_{I_{l,j}}$ as follows, with the warning that  if $q=1$ the couple $(k,q-1)$ is replaced by $(k-1,N-k+1)$ in (\ref{case-in})-(\ref{main-def-V-bis}) --- see Fig. \ref{fig:cases-bis} for a graphical representation of the different cases b), c) d-1) and d-2), below:

\begin{itemize}
\item[a)]
in all the following cases
\begin{itemize}
\item[a-i)] $l\leq k-1$;
\item[a-ii)]  $I_{l,i}\cap I_{k,q}=\emptyset$;
\item[a-iii)]  $ I_{l,i}\cap I_{k,q}\neq\emptyset$ but $l\geq k$ and $I_{k,q} \nsubseteq I_{l,i}$;
\end{itemize}
we define
\begin{equation}\label{case-in}
V^{(k,q)}_{I_{l,i}}:=V^{(k,q-1)}_{I_{l,i}}\,;
\end{equation}
\item[b)]
if $I_{l,i}\equiv I_{k,q}$, we define
\begin{equation}
V^{(k,q)}_{I_{l,i}}:= \sum_{j=1}^{\infty}t^{j-1}(V^{(k,q-1)}_{I_{l,i}})^{diag}_j \,;
\end{equation}
\item[c)]
 if $I_{k,q}\subset I_{l,i}$ and $i, i+l \notin I_{k,q}$, we define
\begin{equation}
V^{(k,q)}_{I_{l,i}}:=\,V^{(k,q-1)}_{I_{l,i}}\,+\sum_{n=1}^{\infty}\frac{1}{n!}\,ad^{n}S_{I_{k,{\color{red}q}}}(V^{(k,q-1)}_{I_{l,i}})\,;
\end{equation}

\item[d)]

\noindent
if $I_{k,q}\subset I_{l,i}$ and either $i$ or $i+l$ belongs to $ I_{k,q}$, we define
\begin{itemize}
\item[ d-1)] if $i$ belongs to $ I_{k,q}$, i.e., $q \equiv i$,  then
\begin{eqnarray}
V^{(k,q)}_{I_{l,i}} &:= &V^{(k,q-1)}_{I_{l,i}}\,+\sum_{j=0}^{k}\sum_{n=1}^{\infty}\frac{1}{n!}\,ad^{n}S_{I_{k,i}}(V^{(k,q-1)}_{I_{l-j,i+j}})\,; \label{main-def-V}
\end{eqnarray}

\item[ d-2)] if $i+l$ belongs to $ I_{k,q}$, i.e., $q +k\equiv i+l$ that means $q\equiv i+l-k$,  then
\begin{eqnarray}
V^{(k,q)}_{I_{l,i}} &:= &V^{(k,q-1)}_{I_{l,i}}\,+\sum_{j=0}^{k}\sum_{n=1}^{\infty}\frac{1}{n!}\,ad^{n}S_{I_{k,i+l-k}}(V^{(k,q-1)}_{I_{l-j,i}})\,. \label{main-def-V-bis}
\end{eqnarray}
\end{itemize}
Notice that  in both cases, d-1) and d-2), the elements of the sets $\{I_{l-j,i+j}\}_{j=1}^{k}$ and $\{I_{l-j,i}\}_{j=1}^{k}$, respectively,  are all the intervals, $\mathscr{I}$, such that $\mathscr{I} \cap I_{k,q} \neq \emptyset $,  $\mathscr{I} \nsubseteq I_{k,q}$,  $I_{k,q}\nsubseteq \mathscr{I} $, and   $\mathscr{I} \cup I_{k,q}\equiv I_{l,i}$.
\end{itemize}
\end{defn}

\begin{rem}
Notice that, according to  Definition \ref{def-interections}:
\begin{itemize}
\item if $(k', q')\succ (l,i)$  then
\begin{equation}
V^{(k',q')}_{I_{l,i}}=V^{(l,i)}_{I_{l,i}}\,,
\end{equation}
since the occurrences in cases b), c), d-1), and d-2) are excluded;
\item
 for $k\geq 1$ and all  allowed choices of $q$, 
\begin{equation}
V^{(k,q)}_{I_{0,i}}=H_i\,
\end{equation}
due to a-i).
\end{itemize}
\end{rem}

\begin{rem}\label{algorithm}Though our analysis of a), b), c), d-1), and d-2) is essentially identical to the corresponding one carried out in the study of quantum chains with bounded interactions (see \cite{FP}), the algorithm is used in Theorem \ref{th-norms} in a different way, since we cannot exploit the unitarity of $e^{S_{I_{k,i}}}$ for the control of the potentials $V^{(k,q)}_{I_{l,i}}$ in terms of their counterparts at the previous step $(k,q-1)$ . Indeed, in contrast to the bounded case treated in \cite{FP},  we use a series expansion of $e^{S_{I_{k,i}}}V^{(k,q-1)}_{I_{l,i}}e^{-S_{I_{k,i}}}$ in cases c), d-1), and d-2).
\end{rem}

By including all potentials\footnote{Recall that $V_{I_{0,i}}^{(0,N)}:=H_i$ and $V_{I_{0,i}}^{(k,q)}$ will coincide with $V_{I_{0,i}}^{(0,N)}$ for all $(k,q)$.} $V^{(k,q-1)}_{I_{l,i}}$, with $I_{l,i}\subset I_{k,q}$,  we obtain the operator denoted by $G_{I_{k,q}}$. Moreover, by construction of $S_{I_{k,q}}$,
\begin{equation}\label{conjugation}
e^{S_{I_{k,q}}}\,(G_{I_{k,q}}+tV^{(k,q-1)}_{I_{k,q}})\,e^{-S_{I_{k,q}}}=G_{I_{k,q}}+t\sum_{j=1}^{\infty}t^{j-1}(V^{(k,q-1)}_{I_{k,q}})^{diag}_j
\end{equation}
where \emph{``diag''} indicates that the corresponding operator is block-diagonal w.r.t. to the decomposition of the identity into $P^{(-)}_{I_{k,q}}+P^{(+)}_{I_{k,q}}$.

\noindent
\emph{Definition of potentials $V^{(k,q)}_{I_{l,i}}$ with $I_{l,i}\subseteq I_{k,q}$}:
\begin{enumerate}
 \item[i)]
If $I_{l,i}\equiv I_{k,q}$ we set
\begin{equation}\label{diag-term}
V^{(k,q)}_{I_{l,i}\equiv I_{k,q}}:=\sum_{j=1}^{\infty}t^{j-1}(V^{(k,q-1)}_{I_{k,q}})^{diag}_j=e^{S_{I_{k,q}}}\,(\frac{G_{I_{k,q}}}{t}+V^{(k,q-1)}_{I_{k,q}})\,e^{-S_{I_{k,q}}}-\frac{G_{I_{k,q}}}{t}\,.
\end{equation} 
Clearly the operator $V^{(k,q)}_{I_{k,q}}$ acts as the identity outside $\mathcal{H}_{I_{k,q}}$ but in general $$\|V^{(k,q)}_{I_{k,q}}\|_{H^0}\neq \|V^{(k,q-1)}_{I_{k,q}}|_{H^0}\,.$$

\item[ii)]
If $I_{l,i}\subset I_{k,q}$  we have set
\begin{equation}\label{def-int}
V^{(k,q)}_{I_{l,i}}:=V^{(k,q-1)}_{I_{l,i}}\,,
\end{equation}
which is block-diagonal w.r.t. the decomposition of the identity into $P^{(+)}_{I_{k,q}}+P^{(-)}_{I_{k,q}}$, too, as explained in Remark \ref{remark-decomp}. Clearly the operator $V^{(k,q)}_{I_{l,i}}$ acts as the identity outside $\mathcal{H}_{I_{l,i}}$ and $\|V^{(k,q)}_{I_{l,i}}\|_{H^0}=\|V^{(k,q-1)}_{I_{l,i}}\|_{H^0}$\,.
\end{enumerate}
We emphasize that
the net result of the conjugation  of the sum of the operators $V^{(k,q-1)}_{I_{l,i}}$  appearing on the left side of  (\ref{conjugation}) can be re-interpreted as follows:

\noindent
a) The operators $V^{(k,q-1)}_{I_{l,i}}$, with $I_{l,i}\subset I_{k,q}$\,, are kept fixed in step $(k,q-1)\rightarrow (k,q)$; i.e., we define $V^{(k,q)}_{I_{l,i}}:=V^{(k,q-1)}_{I_{l,i}}$. Hence
$$G_{I_{k,q}}=\sum_{i\subset I_{k,q} }H_i+t\sum_{I_{1,i} \subset I_{k,q}} V^{(k, q-1)}_{I_{1,i}}+\dots+t\sum_{I_{k-1,i}\subset I_{k,q}}V^{(k, q-1)}_{I_{k-1,i}}=\sum_{i\subset I_{k,q} }H_i+t\sum_{I_{1,i} \subset I_{k,q}} V^{(k, q)}_{I_{1,i}}+\dots+t\sum_{I_{k-1,i}\subset I_{k,q}}V^{(k, q)}_{I_{k-1,i}}\,;$$

\noindent
b)  the operator $V^{(k,q-1)}_{I_{k,q}}$ is transformed to the operator $$V^{(k,q)}_{I_{k,q}}:=\sum_{j=1}^{\infty}t^{j-1}(V^{(k,q-1)}_{I_{k,q}})^{diag}_j$$  which is block-diagonal, and $$\|V^{(k,q)}_{I_{k,q}}\|_{H^0}\leq 2\|V^{(k,q-1)}_{I_{k,q}}\|_{H^0}\,,$$ as will be shown, assuming that $t>0$ is sufficiently small.

\setcounter{equation}{0}

\section{Block-diagonalization of $K_{N}$ - inductive control of $\|V^{(k,q)}_{I_{r,i}}\|_{H^0}$}\label{norms}
In the next theorem, we estimate the weighted norm $$\|V^{(k,q)}_{I_{r,i}}\|_{H^0}:=\|(H_{I_{r,i}}^0+1)^{-\frac{1}{2}}V^{(k,q)}_{I_{r;i}}(H_{I_{r,i}}^0+1)^{-\frac{1}{2}}\|$$ in terms of the norms $\|V^{(k,q-1)}_{I_{l,j}}\|_{H^0}$, i.e., the (weighted) norms of the potentials at the previous block-diagonalization step. For a fixed interval $I_{r,i}$, the weighted norm of the potential does not change, i.e., $\|V^{(k,q-1)}_{I_{r,i}}\|_{H^0}=\|V^{(k,q)}_{I_{r,i}}\|_{H^0}$, 
in step $(k,q-1) \rightarrow (k,q)$, unless some conditions are fulfilled. To gain some intuition of this fact, the reader is advised to take a look at Fig. 1, (replacing $l$ by $r$). Notice that shifting the interval $I_{k,q}$, with $q\geq 2$, to the left by one site makes it coincide with $I_{k,q-1}$. If $I_{k,q}$ is not contained in $I_{r,i}$   then  $\|V^{(k,q)}_{I_{r,i}}\|_{H^0}=\|V^{(k,q-1)}_{I_{r,i}}\|_{H^0}$. Therefore, in step $(k,q-1) \rightarrow (k,q)$,   a change of the weighted norm, i.e., 
$\|V^{(k,q)}_{I_{r,i}}\|_{H^0}\neq \|V^{(k,q-1)}_{I_{r,i}}\|_{H^0}$, may happen in  at most $r-k+1$ cases, provided $r> k$, and only in one case if $k$ coincides with the length $r$\,; and it never happens if $r<k$.  

\noindent
In the theorem below we estimate the change of the weighted norm of the potentials in the block-diagonalization steps, for each $k$, starting from $k=0$.  In the nontrivial steps described above,  we have to make use of a lower bound on the gap above the ground-state energy in the energy spectrum of the Hamiltonian $G_{I_{k,q}}$. This lower bound follows from estimate (\ref{ass-2}), as explained in Lemma \ref{gap} and Corollary \ref{cor-gap}. We will proceed inductively by showing that, for $t(>0)$ sufficiently small but independent of $r$, $N$, $k$, and $q$, the operator-norm bound in (\ref{ass-2}), at step $(k,q-1)$, $q\geq 2$  (for $q=1$ see the footnote), yields control over the spectral gap of the Hamiltonians $G_{I_{k,q}}$, (see Corollary \ref{cor-gap}), and the latter provides an essential  ingredient for the proof of a bound on the weighted  operator norms of the potentials, according to  (\ref{ass-2}), at the next step  $(k,q)$.

\begin{thm}\label{th-norms}
Assume that the coupling constant $t>0$ is sufficiently small uniformly in $k$, $q$, and $N$, and such that Lemma \ref{unboundedlemmaA3} holds true. Then the Hamiltonians $G_{I_{k,q}}$ are well defined, and
\begin{enumerate}
\item[S1)] for any interval  $I_{r,i}$, with $r\geq 1$, the operator $$(H_{I_{r,i}}^0+1)^{-\frac{1}{2}}V^{(k,q)}_{I_{r,i}}(H_{I_{r,i}}^0+1)^{-\frac{1}{2}}$$ has a norm bounded by $t^{\frac{r-1}{4}}$,
 \item[S2)] $G_{I_{k,q+1}}$ has\footnote{  \label{footnote} Recall the special steps of type $(k,q)\equiv (k, N-k)$ with subsequent step $(k+1,1)$.} a spectral gap that is bounded from below by $\Delta_{I_{k,q+1}}\geq \frac{1}{2} $ above the ground state energy,
where  $G_{I_{k,q}}$ is defined in (\ref{def-G}) for $k\geq 2$,  and $G_{I_{1,q}}:=H_{q}+H_{q+1}$.
\end{enumerate}
\end{thm}

\noindent
\emph{Proof.}

\noindent
The proof is by induction in the diagonalization step $(k,q)$, starting at $(k,q)=(0,N)$, and ending at $(k,q)=(N-1,1)$; notice that \textit{S2)}  is not defined for $(k,q)=(N-1,1)$.  

We shall show that for any interval  $I_{r,i}$, with $r\geq 1$, for $(k,q)\prec (r,i+1)$ and for $(k,q)=(r,i+1)$,
\begin{equation}\label{op-norm}
\|(H_{I_{r,i}}^0+1)^{-\frac{1}{2}}V^{(k,q)}_{I_{r,i}}(H_{I_{r,i}}^0+1)^{-\frac{1}{2}}\|\leq \mathcal{E}_{I_{r,i}}^{(k,q)}
\end{equation}
with
\begin{eqnarray}\label{def-estimation}
\quad & &1\quad\quad\quad\quad\quad\quad\quad\quad\quad\quad\quad\quad\quad\quad\quad\quad\quad\quad\quad\quad\quad \text{if}\,\, k= 0, \quad r=1 \nonumber \\
\quad & &0\quad\quad\quad\quad\quad\quad\quad\quad\quad\quad\quad\quad\quad\quad\quad\quad\quad\quad\quad\quad\quad \text{if}\,\, k= 0, \quad r>1 \nonumber \\
\quad & &2^{\chi_{r-k} (q-i)}\,\quad\quad\quad\quad\quad\quad\quad\quad\quad\quad\quad\quad\quad\quad\quad\quad\quad\quad \text{if}\,\,k\geq1,\quad  r= 1 \nonumber \\
\mathcal{E}_{I_{r,i}}^{(k,q)} &:=\,\{& \quad\quad\quad\quad\quad\quad\quad\quad\quad\quad\quad\quad\quad\quad\quad\quad\quad\quad\label{defE}\\
\quad & & \mathcal{Z}_{I_{r,i}}^{(k,q)}2^{\chi_{r-k} (q-i)}\,t^{\frac{r-1}{3}} \quad\quad\quad\quad\quad\quad\quad\quad\quad\quad\quad\quad\quad\quad\quad \text{if}\,\,k\geq1,\quad r= 2 \quad \nonumber \\
\quad & & \underbrace{\mathcal{Z}_{I_{r,i}}^{(k,q)}}\,\underbrace{\Big\{\prod_{s=1}^{g_r(k)}(1+t^{\frac{s}{4}})^{r-s-1}\Big\}\,(1+t^{\frac{k}{4}})^{f_{r-k}(q-i)}}\,\underbrace{2^{\chi_{r-k} (q-i)}}\,t^{\frac{r-1}{3}}\,\quad \text{if}\,\,k\geq1,\quad r\geq 3\nonumber\\
& &\quad \mathcal{I} \quad\quad\quad\quad\quad\quad\quad \mathcal{II} \quad\quad\quad\quad\quad\quad\quad \mathcal{III} \nonumber
\end{eqnarray}
where the factors labeled by $\mathcal{I}$, $\mathcal{II}$, and $\mathcal{III}$ are explained in detail, below.  By construction, for $(r,i+1) \prec  (k,q)$, the operator $V^{(k,q)}_{I_{r,i}}$ coincides with $V^{(r,i+1)}_{I_{r,i}}$; therefore the operator norm on the l-h-s of (\ref{op-norm}) is bounded by $\mathcal{E}_{I_{r,i}}^{(k,q)}\equiv \mathcal{E}_{I_{r,i}}^{(r,i+1)}$.  Hence in the following we shall only consider the cases $(k,q)\prec (r,i+1)$, or $(k,q)=(r,i+1)$.
\\

\noindent
\emph{Definition of factors $\mathcal{I}$, $\mathcal{II}$, and $\mathcal{III}$  (recall  $1\leq k\leq r$)}
\begin{itemize}
\item
factor $\mathcal{I}$ is connected to the contributions to the norm change due to the mechanisms  described by d-1) and d-2) of Definition \ref{def-interections}, i.e., where the interval $I_{k',q'}$ has an endpoint coinciding with either $i$ or $i+r$ and is contained in $I_{r,i}$; factor $\mathcal{I}$ is defined as  (recall  $1\leq k\leq r$)
\begin{equation}\label{def-calS}
\mathcal{Z}_{I_{r,i}}^{(k,q)}:=\sum_{l=1}^{k-1}\frac{2}{(r-l)^{2}\cdot l^2}+\frac{z_{r-k}(q-i)}{(r-k)^{2}\cdot k^2}
\end{equation}
with
\begin{eqnarray}
 z_{r-k}(q-i):=0 & & \text{if}\quad 1\leq r-k\quad \text{and}\quad  q-i< 0\\
  z_{r-k}(q-i):=1&& \text{if}\quad 1\leq r-k\quad \text{and}\quad 0\leq q-i< r-k\\
 z_{r-k}(q-i):=2& & \text{if}\quad 1\leq r-k\quad \text{and}\quad q-i\geq r-k\,\\
  z_{r-k}(q-i):=0& & \text{if}\quad 0= r-k \,,
 \end{eqnarray}
 and the sum $\sum_{l=1}^{k-1}$ is absent if $k= 1$;
\item
factor $\mathcal{II}$ is connected to the contribution to the norm change due to the mechanism  described in c) of Definition \ref{def-interections}, and the functions $g_r(k)$ and $f_{r-k}(q-i)$ are  (recall $1\leq k\leq  r$)
\begin{eqnarray}
  g_r(k):=k-1&& \label{defg}
 \end{eqnarray}
 and
\begin{eqnarray}
 f_{r-k}(q-i):=0 & & \text{if}\quad 2\leq r-k\quad \text{and}\quad  q-i\leq  0 \label{deff-in}\\
  f_{r-k}(q-i):=q-i&& \text{if}\quad 2\leq r-k\quad \text{and}\quad 1\leq q-i\leq r-k-2\\
 f_{r-k}(q-i):=r-k-1& & \text{if}\quad 2\leq r-k\quad \text{and}\quad q-i\geq r-k-1\,\\
  f_{r-k}(q-i):=0& & \text{if}\quad 1\geq r-k \,,\label{deff-fin}
 \end{eqnarray}
respectively,   and the product $\prod_{s=1}^{g_r(k)}$ is absent if $g_r(k)=0$, i.e., for $k= 1$;
 \item
factor $\mathcal{III}$ is connected to the contribution to the norm due to the mechanism described in b) of Definition \ref{def-interections}, and the function $\chi_{r-k}(q-i)$ is defined as follows (recall $1\leq k\leq  r$)
\begin{eqnarray}
\chi_{r-k}(q-i):=0&& \text{if}\quad 1\leq r-k\\
 \chi_{r-k}(q-i):=0& & \text{if}\quad   0= r-k\quad \text{and}\quad q-i\leq -1\\
 \chi_{r-k}(q-i):=1& & \text{if}\quad   0= r-k\quad \text{and}\quad q-i\geq 0\,.
 \end{eqnarray}
\end{itemize}

\begin{rem} 
 Notice that 
 for $t>0$ sufficiently small, but independent of $r,N\geq 1$, $k\geq 1$, and $q$, we have
\begin{equation}\label{est-E}
\mathcal{E}_{I_{r,i}}^{(k,q)}\leq t^{\frac{r-1}{4}}
\end{equation}
by using 
\begin{eqnarray}
& &\exp \Big\{\ln \Big(\Big\{\prod_{s=1}^{g_r(k)}(1+t^{\frac{s}{4}})^{r-s-1}\Big\}\,(1+t^{\frac{k}{4}})^{f_{r-k}(q-i)}\Big)\Big\}\\
&\leq & \exp \Big\{C\cdot \Big[\sum_{s=1}^{g_r(k)}(r-s-1)t^{\frac{s}{4}}+t^{\frac{k}{4}}f_{r-k}(q-i)\Big]\Big\}\\
&\leq & \exp\Big\{C'\cdot t^{\frac{1}{4}}\cdot r\Big\}
\end{eqnarray}
for universal constants $C,C'$. From the definition in (\ref{defE}),  we can readily derive that $\mathcal{E}_{I_{r,i}}^{(0,q)}\leq t^{\frac{r-1}{4}}$. Combining these two observations we deduce that  (\ref{op-norm})  implies  \textit{S1)}.
\end{rem}

For $(k,q)= (0,N)$, we observe that $K^{(k,q)}_N\equiv K_N$ and $G_{I_{k,q}}$ is not defined, indeed it is not needed since   \textit{S1)} is verified by direct computation,  because by definition
$$\|V_{I_{1,i}}^{(0,N)}\|_{H^0}=\| V_{I_{1,i}}\|_{H^0}=\frac{1}{2}<\mathcal{E}_{I_{1,i}}^{(0,N)}\,,$$
and
$V_{I_{r,i}}^{(0,N)} =0$, for $r\geq 2$. Hence (\ref{op-norm}) and, consequently, S1)  hold in step $(0,N)$ for all $I_{r,i}$.  \textit{S2)} holds trivially since, by definition,  the successor of $(0,N)$ is $(1,1)$ and $G_{I_{1,1}}=H_1+H_2$.  
\\

\noindent
Assume that (\ref{op-norm}) and \textit{S2)} hold for all  steps $(k',q')$ with $(k',q') \prec (k,q)$, and recall that  (\ref{op-norm})  implies  \textit{S1)}. We prove that they then hold at step $(k,q)$. By Remark \ref{self}, \textit{S1)} for $ (k,q-1)$ implies that $G_{I_{k,q}}$ is well defined. Furthermore,  \textit{S1)} and \textit{S2)} for $ (k,q-1)$, through Lemma \ref{unboundedlemmaA3},  imply that $S_{I_{k,q}}$ is well defined.  In the steps described  below it is understood that if $q=1$ the couple $(k,q-1)$ is replaced by $(k-1,N-k+1)$.
\\

\noindent
\emph{Induction step in the proof of  (\ref{op-norm}) and S1)}

\noindent
Starting from Definition  \ref{def-interections} we consider the following cases:
\\

\noindent
\emph{Case $r=1$.}
\\

Let $k>1(=r)$ or $k=1=r$ but $I_{1,i}$ such that $i \neq q$.  Then the possible cases are described in a-i), a-ii),  and a-iii), see Definition  \ref{def-interections}, and we have that
\begin{equation}
\|V^{(k,q)}_{I_{1,i}}\|_{H^0}=\|V^{(k,q-1)}_{I_{1,i}}\|_{H^0}\,.
\end{equation} 
Moreover, according to the definition in  (\ref{def-estimation}),  for $k>1$ or for $k=1$ and  $q>i$ we have $\mathcal{E}_{I_{1,i}}^{(k,q-1)}=\mathcal{E}_{I_{1,i}}^{(k,q)}$;  analogously, for $k=1$ and  $q<i$,  $\mathcal{E}_{I_{1,i}}^{(1,q-1)}=\mathcal{E}_{I_{1,i}}^{(1,q)}$. Hence, in the cases discussed above,  by using the inductive hypothesis  we deduce that the property holds for $(k,q)$ if it holds for $(k,q-1)$.
Let $k=1$ and  assume that the set $I_{1,i}$ coincides with $I_{1,q}$. Then  we refer to case b), see Definition  \ref{def-interections}, and to the inductive hypothesis,   and we  find that
\begin{equation}
\|V^{(1,q\equiv i)}_{I_{1,q\equiv i}}\|_{H^0} \leq 2 \|V^{(1,i-1)}_{I_{1,i}}\|_{H^0}\leq 2\mathcal{E}_{I_{1,i}}^{(1,i-1)}\,,
\end{equation}
where the inequality  $\|V^{(1,q)}_{I_{1,q}}\|_{H^0} \leq 2 \|V^{(1,q-1)}_{I_{1,q}}\|_{H^0}$ holds for $t$ sufficiently small, uniformly in $q$ and $N$,  thanks to  Lemma \ref{unboundedlemmaA3}, which can be applied since we assume S1) and S2) at step $(1,q-1)$. To complete the argument it suffices to observe that  $\mathcal{E}_{I_{1,i}}^{(1,i)}=2\mathcal{E}_{I_{1,i}}^{(1,i-1)}$ according to the definition in (\ref{def-estimation}).
\\

\noindent
\emph{Case $r\geq 3$.}
\\

If $\underline{ (r, i+r-k)\prec (k,q)}$ (i.e., either $k>r$ or $k=r$ and $q>i+r-k=i$),  S1) is trivial since
\begin{equation}
V^{(k,q)}_{I_{r,i}}=V^{(k,q-1)}_{I_{r,i}}\,
\end{equation}  
due to Definition  \ref{def-interections},
and 
\begin{equation}
\mathcal{E}^{(k,q)}_{I_{r,i}}=\mathcal{E}^{(k,q-1)}_{I_{r,i}}\,
\end{equation}  
for $(r, i+r-k)\prec (k,q)$, by definition of $\mathcal{E}^{(k,q)}_{I_{r,i}}$. 
\\

If $\underline{(k,q)= (r, i)}$ then we can apply Lemma \ref{unboundedlemmaA3} and estimate
\begin{equation}
\|V^{(k,q\equiv i)}_{I_{r\equiv k,i}}\|_{H^0} \leq 2 \|V^{(k,i-1)}_{I_{r\equiv k,i}}\|_{H^0}\,.
\end{equation}
Hence, by using the inductive hypothesis for $(k,q)=(r, i-1)$, namely  
$$ \|V^{(k,i-1)}_{I_{r\equiv k,i}}\|_{H^0}\leq \mathcal{E}^{(k,i-1)}_{I_{r\equiv k,i}}=\mathcal{Z}_{I_{r,i}}^{(k,q)}\Big\{\prod_{s=1}^{g_r(k)}(1+t^{\frac{s}{4}})^{r-s-1}\Big\}\,(1+t^{\frac{k}{4}})^{f_{r-k}(q-i)}\,t^{\frac{r-1}{3}}\,\Big|_{k\equiv r, q\equiv i-1},$$  we get
\begin{eqnarray}
\|V^{(k,q\equiv i)}_{I_{r\equiv k,i}}\|_{H^0} &\leq&2 \|V^{(k,i-1)}_{I_{r\equiv k,i}}\|_{H^0}\\
& \leq &\mathcal{Z}_{I_{r,i}}^{(k,q)}\Big\{\prod_{s=1}^{g_r(k)}(1+t^{\frac{s}{4}})^{r-s-1}\Big\}\,(1+t^{\frac{k}{4}})^{f_{r-k}(q-i)}\,2^{\chi_{r-k} (q-i)}\,t^{\frac{r-1}{3}}\,\Big|_{r\equiv k,q\equiv i}\nonumber\\
& =&\mathcal{E}^{(k,i)}_{I_{k\equiv r;i}}\,,
\end{eqnarray}
and the property holds for $(k,q)=(r,i)$.
\\

If \underline{$(k,q)\equiv (r,q)$ with $q<i$} the property is trivially valid, because
\begin{equation}
V^{(k,q)}_{I_{r,i}}=V^{(k,q-1)}_{I_{r,i}}\,,\quad \mathcal{E}^{(k,q)}_{I_{r,i}}=\mathcal{E}^{(k,q-1)}_{I_{r,i}},
\end{equation}  
according to Definition  \ref{def-interections} and (\ref{def-estimation}), respectively. 

\noindent
Likewise, for \underline{$(k,q)\equiv (r-1,q)$ with $q\geq i+2$}, the property holds\,. 
\\

For $\underline{(k,q)\equiv (r-1, i+1)}$, we observe that, using case d-2), see  (\ref{main-def-V-bis}),  we derive the estimate
\begin{equation}\label{arg-bis}
\|V^{(k,q)}_{I_{r,i}}\|_{H^0}\leq \|V^{(k,q-1)}_{I_{r,i}}\|_{H^0}+\Big\{\sum_{j=0}^{k}\sum_{n=1}^{\infty}\frac{1}{n!}\,\|(H_{I_{r,i}}^0+1)^{-\frac{1}{2}}\, ad^{n}S_{I_{k,q}}(V^{(k,q-1)}_{I_{r-j,i}})\,(H_{I_{r,i}}^0+1)^{-\frac{1}{2}}\|\,\,.
\end{equation}

\noindent
In order to control
\begin{equation}
\|\sum_{n=1}^{\infty}\frac{1}{n!}\,(H_{I_{r,i}}^0+1)^{-\frac{1}{2}}\,ad^{n}S_{I_{k,q}}(V^{(k, q-1)}_{I_{r-j,i}})\,(H_{I_{r,i}}^0+1)^{-\frac{1}{2}}\|,
\end{equation}
we have to study terms of the type
\begin{equation}
(H_{I_{r,i}}^0+1)^{-\frac{1}{2}}\,S_{I_{k,q}}\dots S_{I_{k,q}}V^{(k,q-1)}_{I_{r-j,i}}S_{I_{k,q}}\dots S_{I_{k,q}}\,(H_{I_{r,i}}^0+1)^{-\frac{1}{2}},
\end{equation}
which we re-write as
\begin{equation}
(H_{I_{r,i}}^0+1)^{-\frac{1}{2}}\,S_{I_{k, q}}\dots S_{I_{k,q}}(H_{I_{r-j,i}}^0+1)^{\frac{1}{2}}(H_{I_{r-j,i}}^0+1)^{-\frac{1}{2}}V^{(k,q-1)}_{I_{r-j,i}}S_{I_{k,q}}\dots S_{I_{k,q}}\,(H_{I_{r,i}}^0+1)^{-\frac{1}{2}}\,.
\end{equation}
Let us show how to bound
\begin{equation}
(H_{I_{r,i}}^0+1)^{-\frac{1}{2}}\,S_{I_{k,q}}\dots S_{I_{k,q}}(H_{I_{r-j,i}}^0+1)^{\frac{1}{2}}\,.
\end{equation}
We insert $\charf=(H_{I_{r-j, i}\setminus I_{k; q}}^0+1)^{\frac{1}{2}}(H_{I_{r-j,i}\setminus I_{k,q}}^0+1)^{-\frac{1}{2}}$ and use that $[H_{I_{r-j, i}\setminus I_{k,q}}^0\,,\,S_{I_{k,q}}]=0$,  which holds since the two supports, $I_{r-j, i}\setminus I_{k,q}$ and $I_{k,q}$, are nonoverlapping by construction. Thus
\begin{eqnarray}
& &(H_{I_{r,i}}^0+1)^{-\frac{1}{2}}(H_{I_{r-j, i}\setminus I_{k,q}}^0+1)^{\frac{1}{2}}(H_{I_{r-j, i}\setminus I_{k,q}}^0+1)^{-\frac{1}{2}}\,S_{I_{k, q}}\dots S_{I_{k,q}}(H_{I_{r-j, i}}^0+1)^{\frac{1}{2}}\nonumber \\
&=&(H_{I_{r,i}}^0+1)^{-\frac{1}{2}}(H_{I_{r-j,i}\setminus I_{k,q}}^0+1)^{\frac{1}{2}}\,S_{I_{k, q}}\dots S_{I_{k,q}}(H_{I_{r-j, i}\setminus I_{k,q}}^0+1)^{-\frac{1}{2}}(H_{I_{r-j, i}}^0+1)^{\frac{1}{2}}\,.\nonumber
\end{eqnarray}
Next we make  use of
\begin{itemize}
\item
the results in Lemma \ref{unboundedlemmaA3} 
\begin{equation}\label{bound-S}
\|S_{I_{k, q}}\|\leq At\,
 \| V^{(k,q-1)}_{I_{k,q}}\|_{H_0}\,,
\end{equation}
\begin{equation}
\|S_{I_{k,q}}(H^0_{I_{k,q}}+1)^{\frac{1}{2}}\|= \|(H^0_{I_{k,q}}+1)^{\frac{1}{2}}S_{I_{k,q}}\|\leq Bt\,
 \| V^{(k,q-1)}_{I_{k,q}}\|_{H_0}\,,
\end{equation}
which follow from the inductive hypotheses for S1) and  S2);
\item
the operator norm bounds
\begin{equation}
\|(H_{I_{r,i}}^0+1)^{-\frac{1}{2}}(H_{I_{r-j, i}\setminus I_{k,q}}^0+1)^{\frac{1}{2}}\|\leq 1\,,
\end{equation}
\begin{equation}
\|(H^0_{I_{k,q}}+1)^{-\frac{1}{2}}(H_{I_{r-j, i}\setminus I_{k,q}}^0+1)^{-\frac{1}{2}}(H_{I_{r-j,i}}^0+1)^{\frac{1}{2}}\|\leq 1
\end{equation}
which follow  from the spectral theorem for commuting operators and from the inclusion $I_{r-j, i}\subset I_{r,i}$.
\end{itemize}
Hence, combining the previous estimates with the inductive hypothesis on
\begin{equation}
\|(H_{I_{r-j, i}}^0+1)^{-\frac{1}{2}}V^{(k,q-1)}_{I_{r-j,i}}(H_{I_{r-j, i}}^0+1)^{-\frac{1}{2}}\|\,,
\end{equation}
and with (\ref{est-E}), we finally conclude that
\begin{equation}\label{estimate-ad}
\|\sum_{n=1}^{\infty}\frac{1}{n!}\,\|(H_{I_{r,i}}^0+1)^{-\frac{1}{2}}\,ad^{n}S_{I_{k,q}}(V^{(k, q-1)}_{I_{r-j,i}})\,(H_{I_{r,i}}^0+1)^{-\frac{1}{2}}\|
\leq C\cdot t\cdot \|V^{(k, q-1)}_{I_{r-j,i}}\|_{H^0}\cdot \|V^{(k, q-1)}_{I_{k,q}}\|_{H^0}
\end{equation}
for $t>0$ sufficiently small but uniform in $N, r, i, k$, and $q$, with $C$ a universal constant. 

\noindent
We recall the inductive hypothesis for the two norms on the r-h-s of (\ref{estimate-ad}). Though we are studying the case $k=r-1$, in the following we make some observations that are useful for general $k$, $k\leq r-1$.  

\noindent
By the definition in (\ref{defE}) and the comment thereafter, supposing that $k< r-j$ or $k= r-j$ and $q-1\leq i+1$, we have that
\begin{eqnarray}
\|V^{(k, q-1)}_{I_{r-j,i}}\|_{H^0}
&\leq &\mathcal{Z}_{I_{r-j,i}}^{(k,q-1)}\Big\{\prod_{s=1}^{g_{r-j}(k)}(1+t^{\frac{s}{4}})^{r-j-s-1}\Big\}\,(1+t^{\frac{k}{4}})^{f_{r-j-k}(q-1-i)}\,2^{\chi_{r-j-k} (q-1-i)}\,t^{\frac{r-j-1}{3}}\,,\quad\quad \label{inductive-1}
\end{eqnarray}
otherwise, i.e., for   $k> r-j$ or $k= r-j$ and $q-1> i+1$,
\begin{eqnarray}
\|V^{(k, q-1)}_{I_{r-j,i}}\|_{H^0}
&\leq &\Big\{\mathcal{Z}_{I_{r-j,i}}^{(k',q'-1)}\Big\{\prod_{s=1}^{g_{r-j}(k')}(1+t^{\frac{s}{4}})^{r-j-s-1}\Big\}\,(1+t^{\frac{k'}{4}})^{f_{r-j-k'}(q'-1-i)}\,2^{\chi_{r-j-k'} (q'-1-i)}\,t^{\frac{r-j-1}{3}}\Big\}|_{k'\equiv r-j\,,\, q'\equiv i+2} \nonumber \\
&=&\mathcal{Z}_{I_{r-j,i}}^{(r-j,i+1)}\Big\{\prod_{s=1}^{g_{r-j}(r-j)}(1+t^{\frac{s}{4}})^{r-j-s-1}\Big\}\,(1+t^{\frac{r-j}{4}})^{f_{0}(1)}\,2^{\chi_{0} (1)}\,t^{\frac{r-j-1}{3}} \,. \label{inductive-2}
\end{eqnarray}
Analogoulsy, we can write
\begin{eqnarray}
\|V^{(k, q-1)}_{I_{k,q}}\|_{H^0}
&\leq & \mathcal{Z}_{I_{k,q}}^{(k,q-1)}\Big\{\prod_{s=1}^{g_{k}(k)}(1+t^{\frac{s}{4}})^{k-s-1}\Big\}\,(1+t^{\frac{k}{4}})^{f_{k-k}(q-1-q)}\,2^{\chi_{k-k}(q-1-q)}\,t^{\frac{k-1}{3}}\,\nonumber \\
&=& \mathcal{Z}_{I_{k,q}}^{(k,q-1)}\Big\{\prod_{s=1}^{g_{k}(k)}(1+t^{\frac{s}{4}})^{k-s-1}\Big\}\,t^{\frac{k-1}{3}}
\end{eqnarray}
since $f_{k-k}(-1)=0$ and $\chi_{k-k}(q-1-q)=0$.

\noindent
From the definition in (\ref{defg}), we notice that
 for $k\leq r-1$
$$g_{k}(k) \leq g_{r}(k)\,,$$
for $k< r-j$ $$g_{r-j}(k)\leq g_{r}(k)\,,$$
and for $r-j\leq k \leq r-1$ $$g_{r-j}(r-j)\leq g_{r}(k)\,.$$
Hence, for $t>0$ sufficiently small, and for  $k< r-j$,
\begin{eqnarray}
& &\Big\{\prod_{s=1}^{g_{r-j}(k)}(1+t^{\frac{s}{4}})^{r-j-s-1}\Big\}\Big\{\prod_{s=1}^{g_{k}(k)}(1+t^{\frac{s}{4}})^{k-s-1}\Big\}\\
&\leq &\Big\{\prod_{s=1}^{g_{r}(k)}(1+t^{\frac{s}{4}})^{k-j-s-1}\Big\}\Big\{\prod_{s=1}^{g_{r}(k)}(1+t^{\frac{s}{4}})^{r-s-1}\Big\}\\
&\leq &e^{C'\cdot (k-j)\cdot \sum_{s=1}^{g_{r}(k)}\,t^{\frac{s}{4}}}\Big\{\prod_{s=1}^{g_{r}(k)}(1+t^{\frac{s}{4}})^{r-s-1}\Big\}\\
&\leq &e^{C\cdot (k-j)\,t^{\frac{1}{4}}}\Big\{\prod_{s=1}^{g_{r}(k)}(1+t^{\frac{s}{4}})^{r-s-1}\Big\}\,,
\end{eqnarray}
where $C'$ and $C$ are universal constants. An analogous estimate holds for $r-j\leq  k \leq r-1$.
Furthermore, from the definition in (\ref{deff-in})-(\ref{deff-fin}), and taking into account (\ref{inductive-1}) and (\ref{inductive-2}), we derive the bounds
$$
f_{r-j-k}(q-1-i)\leq f_{r-k}(q-1-i)\,
$$
 if $k\leq r-j$,
and 
$$
0=f_{0}(1)\leq f_{r-k}(q-i-1)\,
$$
if $r-j<k\leq r-1$.
We also recall
that $$0\leq \chi_{r-j-k'} (q'-1-i)\leq 1,$$ for all admissible values of $k'$ and $q'$, and
$$
\mathcal{Z}_{I_{r,i}}^{(k,q)}\leq \frac{C}{r^2}\,,
$$
for some universal constant $C$; (see the definition in (\ref{def-calS})).


\noindent
Hence, using that $j\leq k$, we can prove the estimate
\begin{eqnarray}
& &\|V^{(k, q-1)}_{I_{r-j,i}}\|_{H^0}\cdot \|V^{(k, q-1)}_{I_{k,q}}\|_{H^0}\\
&\leq &\frac{2C^2}{(r-k)^2\cdot k^2}\cdot t^{-\frac{1}{3}}[e^{C\cdot (k-j)\,t^{\frac{1}{4}}}\,t^{\frac{k-j}{3}}]\,\Big\{\prod_{s=1}^{g_r(k)}(1+t^{\frac{s}{4}})^{r-s-1}\Big\}\,(1+t^{\frac{k}{4}})^{f_{r-k}(q-1-i)}\,2^{\chi_{r-k} (q-1-i)}\,t^{\frac{r-1}{3}}\nonumber \\
&=&\frac{2C^2}{(r-k)^2\cdot k^2}\cdot t^{-\frac{1}{3}}[e^{C\cdot (k-j)\,t^{\frac{1}{4}}}\,t^{\frac{k-j}{3}}]\,\check{\mathcal{E}}_{I_{r,i}}^{(k,q-1)}\,, \label{prod-Veff}
\end{eqnarray}
where
\begin{equation}\label{def-Echeck}
\check{\mathcal{E}}_{I_{r,i}}^{(k,q-1)}:=\Big\{\prod_{s=1}^{g_r(k)}(1+t^{\frac{s}{4}})^{r-s-1}\Big\}\,(1+t^{\frac{k}{4}})^{f_{r-k}(q-1-i)}\,2^{\chi_{r-k} (q-1-i)}\,t^{\frac{r-1}{3}}=\frac{\mathcal{E}_{I_{r,i}}^{(k,q-1)}}{\mathcal{Z}_{I_{r,i}}^{(k,q-1)}}\,.
\end{equation}
We recall that
\begin{eqnarray}\label{arg-bis}
\|V^{(k,q)}_{I_{r,i}}\|_{H^0}&\leq & \|V^{(k,q-1)}_{I_{r,i}}\|_{H^0}+\Big\{\sum_{j=0}^{k}\sum_{n=1}^{\infty}\frac{1}{n!}\,\|(H_{I_{r,i}}^0+1)^{-\frac{1}{2}}\, ad^{n}S_{I_{k,q}}(V^{(k,q-1)}_{I_{r-j;i}})\,(H_{I_{r,i}}^0+1)^{-\frac{1}{2}}\|\quad\quad\\
&\leq&\|V^{(k,q-1)}_{I_{r,i}}\|_{H^0}+ C\cdot t\cdot \sum_{j=0}^{k}\|V^{(k, q-1)}_{I_{r-j,i}}\|_{H^0}\cdot \|V^{(k, q-1)}_{I_{k,q}}\|_{H^0}\,.
\end{eqnarray}
Therefore, by exploiting (\ref{prod-Veff}) and using  the definition in (\ref{def-Echeck}),  for $t>0$ sufficiently small but uniformly in $k,q$, $r$, and $i$, we find that
\begin{eqnarray}
\|V^{(k,q)}_{I_{r,i}}\|_{H^0}&\leq&\mathcal{Z}_{I_{r,i}}^{(k,q-1)}\check{\mathcal{E}}_{I_{r,i}}^{(k,q-1)}+\frac{2C^2}{(r-k)^2\cdot k^2}\cdot t^{\frac{1}{3}}\cdot \Big\{\sum_{j=0}^{k}[e^{C\cdot (k-j)\,t^{\frac{1}{4}}}\,t^{\frac{k-j}{3}}]\Big\}\,\check{\mathcal{E}}_{I_{r,i}}^{(k,q-1)}\\
&\leq &\mathcal{Z}_{I_{r,i}}^{(k,q-1)}\check{\mathcal{E}}_{I_{r,i}}^{(k,q-1)}+\frac{1}{(r-k)^2\cdot k^2}\check{\mathcal{E}}_{I_{r,i}}^{(k,q-1)}\\
&\leq  &(\mathcal{Z}_{I_{r,i}}^{(k,q-1)}+\frac{1}{(r-k)^2\cdot k^2})\,\check{\mathcal{E}}_{I_{r,i}}^{(k,q-1)}\\
&=&\mathcal{Z}_{I_{r,i}}^{(k,q)}\,\check{\mathcal{E}}_{I_{r,i}}^{(k,q-1)}\\
&\leq  &\mathcal{E}_{I_{r,i}}^{(k,q)}\,.
\end{eqnarray}
If \underline{$k=r-1$ and $q =i$} we proceed in a similar way, exploiting d-1) in Definition  \ref{def-interections}.
\\

For $\underline{k\leq r-2}$,  besides the mechanism already shown that involves an interval $I_{k,q}$ with one of the endpoints coinciding with an endpoint of $I_{r,i}$,  we have to show that the step from $q-1$ to $q$ holds for \emph{inner} intervals, i.e., for intervals $I_{k,q}$ with $i+1\leq q \leq r-k+i-1$. Hence we have to study the r-h-s of
\begin{equation}
\|V^{(k,q)}_{I_{r,i}}\|_{H^0}\leq \|V^{(k,q-1)}_{I_{r,i}}\|_{H^0}+\sum_{n=1}^{\infty}\frac{1}{n!}\,\|(H_{I_{r,i}}^0+1)^{-\frac{1}{2}}\, ad^{n}S_{I_{k,q}}(V^{(k,q-1)}_{I_{r;i}})\,(H_{I_{r,i}}^0+1)^{-\frac{1}{2}}\|,\,\,
\end{equation}
which, for $t$ sufficiently small but independent of $N$, $k$, $q$, $r$,  and $i$, we can estimate as follows:
\begin{eqnarray}\label{arg-bis}
& & \|V^{(k,q-1)}_{I_{r,i}}\|_{H^0}+\sum_{n=1}^{\infty}\frac{1}{n!}\,\|(H_{I_{r,i}}^0+1)^{-\frac{1}{2}}\, ad^{n}S_{I_{k,q}}(V^{(k,q-1)}_{I_{r;i}})\,(H_{I_{r,i}}^0+1)^{-\frac{1}{2}}\|\,\\
&\leq&\|V^{(k,q-1)}_{I_{r,i}}\|_{H^0}+ C\cdot t\cdot \|V^{(k, q-1)}_{I_{k,q}}\|_{H^0}\cdot  \|V^{(k, q-1)}_{I_{r,i}}\|_{H^0}\\
&\leq &\mathcal{E}_{I_{r,i}}^{(k,q-1)}+C\cdot  t\cdot t^{\frac{k-1}{4}}\,\mathcal{E}_{I_{r,i}}^{(k,q-1)}\\
&\leq &(1+t^{\frac{k}{4}})\mathcal{E}_{I_{r,i}}^{(k,q-1)}\\
&= &\mathcal{E}_{I_{r,i}}^{(k,q)}
\end{eqnarray}
\\

\noindent
\emph{Case $r=2$.}
\\

This case is similar to the case where $r\geq 3$ but is actually simpler, since $k(\geq 1)$ cannot be less than or equal to $r-2$, which implies that the corresponding $\mathcal{E}_{I_{r,i}}^{(k,q)}$ does not contain factor $\mathcal{II}$; (see (\ref{def-estimation})).
\\

\noindent
\emph{Induction step to prove S2)}

\noindent
Having proven \textit{S1)}, we can apply Lemma \ref{gap} and Corollary \ref{cor-gap}. Hence, \textit{S2)} holds for $t>0$ sufficiently small, but independent of $N$, $k$, and $q$. \qed

In the next theorem we prove that Definition \ref{def-interections} yields operators $V_{I_{l,j}}^{(k,q)}$ consistent with identity (\ref{def-KN}) between the Hamiltonian $K_{N}^{(k,q)}$ given in (\ref{def-transf-ham})-(\ref{def-transf-ham-bis}) and the conjugation of $K_{N}^{(k,q-1)}$ using $e^{S_{I_{k,q}}}$.
\begin{thm}\label{th-potentials}
Under the assumptions of Theorem \ref{th-norms},
the operator $K_{N}^{(k,q)}$, with $k\geq1$ and $q\geq 2$, defined in (\ref{def-transf-ham})-(\ref{def-transf-ham-bis}) is self-adjoint on the domain $e^{S_{I_{k,q}}}D(K_N^{(k,q-1)})$ and coincides with $e^{S_{I_{k,q}}}\,K_N^{(k,q-1)}\,e^{-S_{I_{k,q}}}$. If $q=1$ the statement holds with $(k,q-1)$ replaced by $(k-1, N-k+1)$.
\end{thm}

\noindent
\emph{Proof.}

\noindent
We study the case $q\geq 2$ explicitly; the case $q=1$ can be proven in the same way. First we prove that the identiy claimed in the statement holds formally. 
Indeed, in the expression
\begin{eqnarray}
e^{S_{I_{k,q}}}\,K_N^{(k,q-1)}\,e^{-S_{I_{k,q}}}&=&
e^{S_{I_{k,q}}}\,\Big[\sum_{i=1}^{N}H_{i}+t\sum_{i=1}^{N-1}V^{(k,q-1)}_{I_{1,i}}+t\sum_{i=1}^{N-2}V^{(k,q-1)}_{I_{2,i}}+\dots+t\sum_{i=1}^{N-k}V^{(k,q-1)}_{I_{k,i}} \nonumber \\
& &+\sum_{i=1}^{N-k-1}V^{(k,q-1)}_{I_{k+1,i}}+\dots+t\sum_{i=1}^{2}V^{(k,q-1)}_{I_{N-2,i}}+tV^{(k,q-1)}_{I_{N-1,1}}\Big]e^{-S_{I_{k,q}}}\,
\end{eqnarray}
we observe that:

\begin{itemize}
\item
For intervals  $I_{l,i}$ such  that $I_{l,i} \cap I_{k;q}=\emptyset$, 
\begin{equation}
e^{S_{I_{k,q}}}V^{(k,q-1)}_{I_{l,i}}e^{-S_{I_{k,q}}}=V^{(k,q-1)}_{I_{l,i}}=:V^{(k,q)}_{I_{l,i}},
\end{equation}
which follows from a-ii), Definition \ref{def-interections}.
\item
For the terms constituting $G_{I_{k,q}}$ (see definition (\ref{def-G})), we get, after adding $tV^{(k,q-1)}_{I_{k,q}}$, 
\begin{eqnarray}
& &e^{S_{I_{k,q}}}\,(G_{I_{k,q}}+tV^{(k,q-1)}_{I_{k,q}})\,e^{-S_{I_{k,q}}}\\
&= &\sum_{i\subset I_{k,q} }H_i+t\sum_{I_{1,i} \subset I_{k,q}} V^{(k, q-1)}_{I_{1,i}}+\dots+t\sum_{I_{k-1,i}\subset I_{k;q}}V^{(k, q-1)}_{I_{k-1,i}}
+t\sum_{j=1}^{\infty}t^{j-1}(V^{(k,q-1)}_{I_{k,q}})^{diag}_j \nonumber\\
&= &\sum_{i\subset I_{k,q} }H_i+t\sum_{I_{1;i} \subset I_{k,q}} V^{(k, q)}_{I_{1,i}}+\dots+t\sum_{I_{k-1,i}\subset I_{k,q}}V^{(k, q)}_{I_{k-1,i}}\label{local-ham}
+tV^{(k,q)}_{I_{k,q}}\,\,,
\end{eqnarray}
where the first identity is the result of the Lie-Schwinger conjugation and the last identity follows from Definition \ref{def-interections}, cases a-i) and b).
\item
Regarding the terms $V^{(k,q-1)}_{I_{l,i}}$,  with $ I_{k,q} \subset I_{l,i}$ and $i,i+l \notin  I_{k,q}$, the expression
\begin{equation}
e^{S_{I_{k,q}}}\,V^{(k,q-1)}_{I_{l,i}}e^{-S_{I_{k,q}}}
\end{equation}
corresponds to $V^{(k,q)}_{I_{l,i}}$, by Definition \ref{def-interections}, case c).
\item
With regard to the terms $V^{(k,q-1)}_{I_{l,i}}$, with  $I_{l,i}\cap I_{k,q}\neq\emptyset$, but $I_{l,i} \nsubseteq I_{k,q}$ and $I_{k,q} \nsubseteq I_{l,i}$, it follows that
\begin{equation}
e^{S_{I_{k,q}}}\,V^{(k,q-1)}_{I_{l,i}}\,e^{-S_{I_{k,q}}}=V^{(k,q-1)}_{I_{l,i}}+\sum_{n=1}^{\infty}\frac{1}{n!}\,ad^{n}S_{I_{k,q}}(V^{(k,q-1)}_{I_{l,i}}) \,.\label{growth-bis}
\end{equation}
The first term on the right side is $V^{(k,q)}_{I_{l,i}}$ (see cases a-i) and a-iii) in Definition \ref{def-interections}), the second term contributes to $V^{(k,q)}_{I_{r,j}}$, where $I_{r;j} \equiv I_{l,i}\cup I_{k,q}$,  together with further similar terms and with
\begin{equation}\label{extra}
e^{S_{I_{k,q}}}\,V^{(k,q-1)}_{I_{r,j}}\,e^{-S_{I_{k,q}}}\,,
\end{equation}
where the set $I_{r,j}$ has the property that $I_{k,q}\subset I_{r,j}$, and either $j$ or $j+r$ belong to $ I_{k,q}$. We observe that the term in (\ref{extra})  has not been considered in the previous cases and corresponds to the first term plus the summands associated with $j=0$ on the r-h-s of  (\ref{main-def-V})  or the analogous quantity in (\ref{main-def-V-bis}),  where $l$ is replaced by $r$ and $i$ by $j$.
\end{itemize}
Hence we get that at least formally
\begin{eqnarray}
e^{S_{I_{k,q}}}\,K_N^{(k,q-1)}\,e^{-S_{I_{k,q}}}&=&
\sum_{i=1}^{N}H_{i}+t\sum_{i=1}^{N-1}V^{(k,q)}_{I_{1,i}}+t\sum_{i=1}^{N-2}V^{(k,q)}_{I_{2,i}}+\dots+t\sum_{i=1}^{N-k}V^{(k,q)}_{I_{k,i}} \nonumber \\
& &+\sum_{i=1}^{N-k-1}V^{(k,q)}_{I_{k+1,i}}+\dots+t\sum_{i=1}^{2}V^{(k,q)}_{I_{N-2,i}}+tV^{(k,q-1)}_{I_{N-1,1}}\,,\label{ham-identity}
\end{eqnarray}
where the operator on the r-h-s is $K_N^{(k,q)}$, by definition.
Our final goal is to prove that (\ref{ham-identity}) is an identity between two self-adjoint operators.  (As for the l-h-s, $K_N^{(k,q-1)}$ is self-adjoint, by assumption, and $e^{-S_{I_{k,q}}}$ is unitary.) 

\noindent
To show this, we need the following input: 
The domain  $D((H^0_{I_{N-1,1}})^{\frac{1}{2}})$ is invariant under  $e^{S_{I_{k,q}}}$.
Indeed, for any $\varphi \in D((H^0_{I_{N-1,1}})^{\frac{1}{2}})$ and $m\in \mathbb{N}$, we claim that
\begin{eqnarray}
& &\|(H^0_{I_{r,i}})^{\frac{1}{2}}\,(S_{I_{k,q}})^m\varphi\| \label{domain-in}\\
&=&\|(H^0_{I_{r,i}})^{\frac{1}{2}}\,\frac{1}{(H^0_{I_{r,i}\setminus I_{k,q}}+1)^{\frac{1}{2}}}(H^0_{I_{r,i}\setminus I_{k,q}}+1)^{\frac{1}{2}}(S_{I_{k,q}})^m\varphi\| \\
&=&\|(H^0_{I_{r,i}})^{\frac{1}{2}}\,\frac{1}{(H^0_{I_{r,i}\setminus I_{k,q}}+1)^{\frac{1}{2}}}  (S_{I_{k,q}})^m \,(H^0_{I_{r,i}\setminus I_{k,q}}+1)^{\frac{1}{2}}\varphi\|\\
&\leq &\|\frac{(H^0_{I_{r,i}})^{\frac{1}{2}}}{(H^0_{I_{r,i}\setminus I_{k,q}}+1)^{\frac{1}{2}}(H^0_{ I_{k,q}}+1)^{\frac{1}{2}}}\|\,\|(H^0_{I_{k,q}}+1)^{\frac{1}{2}}S_{I_{k,q}}\|\,\|S_{I_{k,q}}\|^{m-1}\,\|(H^0_{I_{r,i}\setminus I_{k,q}}+1)^{\frac{1}{2}}\varphi \| \\
&\leq &C_{\varphi}^m\,, \label{domain-fin}
\end{eqnarray}
for some constant $C_{\varphi}$ depending on $\varphi$, where we have exploited  estimate (\ref{S-Hest}) in Lemma \ref{unboundedlemmaA3}, the spectral theorem for commuting self-adjoint operators, and the assumption that $\varphi \in D((H^0_{I_{N-1,1}})^{\frac{1}{2}})$.

\begin{rem}\label{symmetric}
We observe that  assuming at step $(k,q-1)$ that, for any interval $I_{r,i}$, the operators
\begin{equation}
(H^0_{I_{r,i}}+1)^{-\frac{1}{2}}\,V^{(k,q-1)}_{I_{r;i}}\,(H^0_{I_{r,i}}+1)^{-\frac{1}{2}}
\end{equation}
are symmetric, thanks to Theorem \ref{th-norms} and Lemma \ref{unboundedlemmaA3},  we conclude that the definitions  in  (\ref{case-in})-(\ref{main-def-V-bis})   hold in the sense of symmetric, quadratic forms on the domain $D((H^0_{I_{l,i}})^{\frac{1}{2}})$.
\end{rem}
Next, for $t>0$ small enough, as in Theorem \ref{th-norms}, we conclude that the r-h-s of (\ref{ham-identity}) is a symmetric operator bounded from below on the domain $D(H^0_{I_{N-1,1}})$. Starting from this bound,  and arguing as in the procedure used for $K_N$ in Sect.  \ref{intro-def}, we can define a self-adjoint extension for $K_N^{(k,q)}$ (again denoted by $K_N^{(k,q)}$) with domain $D(K_N^{(k,q)})\supseteq D(H^0_{I_{N-1,1}})$ contained in $D((H^0_{I_{N-1,1}})^{\frac{1}{2}})$.

We shall prove by induction that, for $(0,N)\prec (k,q)$, $K_N^{(k,q)}$ coincides with the self-adjoint operator $e^{S_{I_{k,q}}}\,K_N^{(k,q-1)}\,e^{-S_{I_{k,q}}}$ defined on $e^{S_{I_{k,q}}}D(K_N^{(k,q-1)})$ with the property  \begin{equation}\label{domains-property}
D(H^0_{I_{N-1,1}})\subseteq D(K_N^{(k,q)})=e^{S_{I_{k,q}}}D(K_N^{(k,q-1)})\subseteq D((H^0_{I_{N-1,1}})^{\frac{1}{2}})\,.
\end{equation}

\noindent
\emph{Inductive step}

\noindent
For  $(0,N)\prec (k,q-1)$ we assume that $K_N^{(k,q-1)}\equiv e^{S_{I_{k,q-1}}}\,K_N^{(k,q-2)}\,e^{-S_{I_{k,q-1}}}$
 and~$D(K_N^{(k,q-1)})\subseteq D((H^0_{I_{N-1,1}})^{\frac{1}{2}})$. Then   we deduce from the argument outlined in (\ref{domain-in})-(\ref{domain-fin})  that $e^{S_{I_{k;q}}}D(K_N^{(k,q-1)})\subseteq D((H^0_{I_{N-1,1}})^{\frac{1}{2}})$. Next, using the same type of manipulations and estimates as in the proof of Theorem \ref{th-norms} we derive that the relation in (\ref{ham-identity})  holds as an identity between quadratic forms  in the common domain $D((H^0_{I_{N-1, 1}})^{\frac{1}{2}})$, i.e., on the l-h-s of (\ref{ham-identity}) we can expand the exponential operator and control the series whenever we consider a matrix element with vectors in $D((H^0_{I_{N-1, 1}})^{\frac{1}{2}})$ and then check that they correspond to the analogous matrix element of the terms on the r-h-s. In this operation one has to make sure that  the off-diagonal terms  that cancel each other on the l-h-s of (\ref{local-ham}) are individually  well defined. (This cancellation is indeed the purpose of the conjugation.) Indeed, for any $V^{(k,q-1)}_{I_{k,q}}$ and any $j\geq 1$, and for $\varphi, \psi$ in $D((H^0_{I_{N-1,1}})^{\frac{1}{2}})$, the following matrix elements 
\begin{equation}
\langle [(V^{(k,q-1)}_{I_{k,q}})_j-(V^{(k,q-1)}_{I_{k,q}})^{diag}_j]\,\varphi\,,\, \psi\rangle \,
\end{equation}
are well defined due to (\ref{V-ineq}) and to estimates (\ref{diag-est-in})-(\ref{diag-est-fin}).

Since the two self-adjoint operators induce the same closed quadratic form on the domain $D((H^0_{I_{N-1, 1}})^{\frac{1}{2}})$, they coincide. 
The equality $K_N^{(k,q)} = e^{S_{I_{k,q}}}\,K_N^{(k,q-1)}\,e^{-S_{I_{k,q}}} $ implies the inclusions in (\ref{domains-property}).

\noindent
\emph{First step}

\noindent
Notice that for $(k,q)=(0,N)$  the inclusions $D(H^0_{I_{N-1,1}}) \subseteq D(K_N^{(0,N)})\equiv D(K_N)\subseteq D((H^0_{I_{N-1,1}})^{\frac{1}{2}})$ hold true; see Section \ref{intro-def}.  Hence, by the argument outlined in  (\ref{domain-in})-(\ref{domain-fin}) we get that $e^{S_{I_{1,1}}}D(K_N^{(0,N)})\subseteq D((H^0_{I_{N-1,1}})^{\frac{1}{2}})$ and the rest of the proof is analogous to the \emph{Inductive step}. \qed
\begin{thm}\label{main-res}
Under the assumption that (\ref{gaps}), (\ref{potential}) and (\ref{weighted}) hold, the Hamiltonian $K_{N}$ defined in (\ref{Hamiltonian}) has the following properties: There exists some $t_0 > 0$ such that, for any $t\in \mathbb{R}$ with 
$\vert t \vert < t_0$, and for all $N < \infty$,
\begin{enumerate}
\item[(i)]{ $K_{N}\equiv K_N(t)$ has a unique ground-state; and}
\item[(ii)]{ the energy spectrum of $K_N$ has a strictly positive gap, $\Delta_{N}(t) \geq \frac{1}{2}$, above the ground-state energy.}
\end{enumerate}
\end{thm}

\noindent
\emph{Proof.}
Notice that $K_N^{(N-1,1)} \equiv G_{I_{N-1,1}}+tV^{(N-1,1)}_{I_{N-1,1}}$. We have constructed the unitary conjugation 
$e^{S_{N}(t)}$,  (see (\ref{conjug})), such that  the operator
$$e^{S_{N}(t)}K_{N}(t)e^{-S_{N}(t)}=G_{I_{N-1,1}}+tV^{(N-1,1)}_{I_{N-1,1}}=: \widetilde{K}_{N}(t),$$  
has the properties in (\ref{block-diag}) and (\ref{gapss}), which follow from Theorem \ref{th-norms} and from (\ref{final-eq-1}) and (\ref{final-eq-2}), for $(k,q)=(N-1,1)$, where we also include the block-diagonalized potential $V^{(N-1,1)}_{I_{N-1,1}}$. \qed

\setcounter{equation}{0}
\begin{appendix}
\section{Appendix}\label{app}

\begin{lem}\label{op-ineq-1} For any $1\leq n \leq N$
\begin{equation}\label{main-ineq}
\sum_{i=1}^{n} P^{\perp}_{\Omega_i} \geq \charf -\bigotimes_{i=1}^{n} P_{\Omega_i}=:\,\Big(\bigotimes_{i=1}^{n} P_{\Omega_i}\Big)^{\perp}
\end{equation}
where $P^{\perp}_{\Omega_i}=\charf-P_{\Omega_i}$.

\end{lem}

\noindent
\emph{Proof}

\noindent
This lemma coincides with Lemma A.1 of \cite{FP}, where the reader can find the proof.
\qed

From Lemma \ref{op-ineq-1} we derive the following bound.
\begin{cor}\label{op-ineq-2}
For $i+r\leq N$, we define
\begin{equation}
P^{(+)}_{I_{r,i}}:=\Big(\bigotimes_{k=i}^{i+r}P_{\Omega_{k}}\Big)^{\perp}\,.
\end{equation}
Then, for $1\leq l \leq L \leq N-r$, 
\begin{equation}\label{ineq-inter}
\sum_{i=l}^{L}P^{(+)}_{I_{r,i}}\leq (r+1) \sum_{i=l}^{L+r} P^{\perp}_{\Omega_i} \,.
\end{equation}
\end{cor}

\noindent
\emph{Proof}

\noindent
From Lemma \ref{op-ineq-1} we derive
\begin{equation}\label{first-lemma1.3}
\sum_{j=i}^{i+r} P^{\perp}_{\Omega_{j}}\geq \Big(\bigotimes_{k=i}^{i+r}P_{\Omega_{k}}\Big)^{\perp}\,.
\end{equation}
 By summing the l-h-s of (\ref{first-lemma1.3}) for $i$ from $l$ up to $L$, for each $j$ we get not more than $r+1$ terms of the type
$P^{\perp}_{\Omega_{j}}$
and the inequality in (\ref{ineq-inter}) follows\,. \qed

\begin{lem}\label{formboundedness}
For $t>0$ sufficiently small  as stated in Corollary \ref{cor-gap}, the following bound holds
\begin{equation}
(\Phi, P^{(+)}_{I_{k,q}}(G_{I_{k,q}}-E_{I_{k,q}})P^{(+)}_{I_{k,q}}\Phi)\geq \frac{\Delta_{I_{k,q}}}{2} (\Phi, P^{(+)}_{I_{k,q}}(H^0_{I_{k,q}}+1)P^{(+)}_{I_{k,q}}\Phi) \label{bound-G}
\end{equation}
for any vector $\Phi$ in the domain of $H^0_{I_{k,q}}$, where $\Delta_{I_{k,q}}$ is the lower bound of the spectral gap determined in Corollary \ref{cor-gap}. Consequently, 
\begin{equation}
\left\|\frac{1}{(G_{I_{k,q}}-E_{I_{k,q}})^{\frac{1}{2}}}P^{(+)}_{I_{k,q}} (H^0_{I_{k,q}}+1)^{\frac{1}{2}}\right\|\leq \frac{\sqrt{2}}{\Delta_{I_{k,q}}^{\frac{1}{2}}}\label{op-norm-G}
\end{equation}
and
\begin{equation}
\left\|\frac{1}{(G_{I_{k,q}}-E_{I_{k,q}})}P^{(+)}_{I_{k,q}} (H^0_{I_{k,q}}+1)^{\frac{1}{2}}\right\|\leq \frac{\sqrt{2}}{\Delta_{I_{k,q}}}\,.\label{op-norm-G-2}
\end{equation}
\end{lem}

\noindent
\emph{Proof.}

\noindent
The proof of (\ref{bound-G}) follows from inequality (\ref{final-eq-1}) stated in Lemma \ref{gap}.
Regarding the operator norm in (\ref{op-norm-G}), we estimate
\begin{equation}\label{squared-norm}
\left\|(H^0_{I_{k,q}}+1)^{\frac{1}{2}}P^{(+)}_{I_{k,q}} \frac{1}{(G_{I_{k,q}}-E_{I_{k,q}})^{\frac{1}{2}}}\Psi\right\|^2
\end{equation} 
for vectors $\Psi$ of the form 
$\frac{  (G_{I_{k,q}}-E_{I_{k,q}})^{\frac{1}{2}}\Phi}{\|(G_{I_{k,q}}-E_{I_{k,q}})^{\frac{1}{2}}\Phi \|}$, where
$\Phi$ is in the domain of $G_{I_{k,q}}P^{(+)}_{I_{k,q}}=P^{(+)}_{I_{k,q}}G_{I_{k,q}}P^{(+)}_{I_{k,q}}$.
The squared  norm in (\ref{squared-norm}) is seen to coincide with 
\begin{equation}
\frac{(\Phi, (H^0_{I_{k,q}}+1)\Phi)}{(\Phi, (G_{I_{k,q}}-E_{I_{k,q}})\Phi)}\leq \frac{2}{\Delta_{I_{k,q}}}\,,
\end{equation}
where the inequality above corresponds to  (\ref{bound-G}). 
The operator norm in (\ref{op-norm-G-2}) follows from (\ref{op-norm-G}) and Corollary \ref{gap} which implies  $$\|\frac{1}{P^{(+)}_{I_{k,q}}(G_{I_{k,q}}-E_{I_{k,q}})^{\frac{1}{2}}P^{(+)}_{I_{k,q}}}\| \leq \frac{1}{\Delta_{I_{k,q}}^{\frac{1}{2}}}.$$ \qed

\begin{lem}\label{unboundedlemmaA3}
Assume that $t>0$ is sufficiently small,  $\|V^{(k,q-1)}_{I_{r,i}}\|_{H_0} \leq t^{\frac{r-1}{4}}$, and $\Delta_{I_{k,q}}\geq \frac{1}{2}$. Then, for arbitrary $N$, $k\geq 1$, and $q\geq 2$, the inequalities
\begin{equation}\label{bound-V}
\|V^{(k,q)}_{I_{k,q}}\|_{H_0}\leq 2\|V^{(k,q-1)}_{I_{k,q}}\|_{H_0}\,
\end{equation}
\begin{equation}\label{bound-S}
\|S_{I_{k,q}}\|\leq At\,
 \| V^{(k,q-1)}_{I_{k,q}}\|_{H_0}
\end{equation}
\begin{equation}\label{S-Hest}
\|S_{I_{k,q}}(H^0_{I_{k,q}}+1)^{\frac{1}{2}}\|=\|(H^0_{I_{k,q}}+1)^{\frac{1}{2}}S_{I_{k,q}}\| \leq Bt\,
 \| V^{(k,q-1)}_{I_{k,q}}\|_{H_0}
\end{equation}
hold true for universal constants $A$ and $B$. For $q=1$,   $ V^{(k,q-1)}_{I_{k,q}}$  is replaced by $V^{(k-1,N-k+1)}_{I_{k,q}}$ in the right side of (\ref{bound-V}) and (\ref{bound-S}).
\end{lem}

\noindent
\emph{Proof}

\noindent
In the following we assume $q\geq 2$; if $q=1$ an analogous proof holds. We recall that 
\begin{equation}
V^{(k,q)}_{I_{k,q}}:= \sum_{j=1}^{\infty}t^{j-1}(V^{(k,q-1)}_{I_{k,q}})^{diag}_j \,
\end{equation}
and
\begin{equation}
S_{I_{k,q}}:=\sum_{j=1}^{\infty}t^j(S_{I_{k,q}})_j\
\end{equation}
with $$(V^{(k,q-1)}_{I_{k,q}})_1=V^{(k,q-1)}_{I_{k,q}}\,,$$ 
and, for $j\geq 2$,
\begin{eqnarray}
& &(V^{(k,q-1)}_{I_{k,q}})_j\,:=\label{formula-v_j-bis}\\
& &\sum_{p\geq 2, r_1\geq 1 \dots, r_p\geq 1\, ; \, r_1+\dots+r_p=j}\frac{1}{p!}\text{ad}\,(S_{I_{k;q}})_{r_1}\Big(\text{ad}\,(S_{I_{k,q}})_{r_2}\dots (\text{ad}\,(S_{I_{k,q}})_{r_p}(G_{I_{k,q}}) \Big)\\
& &+\sum_{p\geq 1, r_1\geq 1 \dots, r_p\geq 1\, ; \, r_1+\dots+r_p=j-1}\frac{1}{p!}\text{ad}\,(S_{I_{k,q}})_{r_1}\Big(\text{ad}\,(S_{I_{k,q}})_{r_2}\dots (\text{ad}\,(S_{I_{k,q}})_{r_p}(V^{(k,q-1)}_{I_{k,q}}) \Big)\,,\quad\quad\quad\quad\,.
\end{eqnarray}
and
\begin{equation}
(S_{I_{k,q}})_j:=ad^{-1}\,G_{I_{k,q}}\,((V^{(k,q-1)}_{I_{k,q}})^{od}_j)=\frac{1}{G_{I_{k,q}}-E_{I_{k,q}}}P^{(+)}_{I_{k,q}}\,(V^{(k,q-1)}_{I_{k,q}})_j\,P^{(-)}_{I_{k,q}}-h.c.\,.
\end{equation}
where $j\geq 1$.

\noindent
From the lines above we derive
\begin{eqnarray}
\text{ad}\,(S_{I_{k,q}})_{r_p}(G_{I_{k,q}})
&=&\text{ad}\,(S_{I_{k,q}})_{r_p}(G_{I_{k,q}}-E_{I_{k,q}})\nonumber \\
&=&\,[\frac{1}{G_{I_{k,q}}-E_{I_{k,q}}}P^{(+)}_{I_{k,q}}\,(V^{(k, q-1)}_{I_{k,q}})_{r_p}\,P^{(-)}_{I_{k,q}}\,,\,G_{I_{k,q}}-E_{I_{k,q}}]+h.c.\\
&=&-P^{(+)}_{I_{k,q}}\,(V^{(k,q-1)}_{I_{k,q}})_{r_p}\,P^{(-)}_{I_{k,q}}-P^{(-)}_{I_{k,q}}\,(V^{(k, q-1)}_{I_{k,q}})_{r_p}\,P^{(+)}_{I_{k,q}}\,.
\end{eqnarray}

\noindent
We start by showing the following inequality
\begin{equation}\label{Snorm}
\|(S_{I_{k,q}})_j\|\leq \frac{2\sqrt{2}}{\Delta_{I_{k,q}}} \| (V_{I_{k,q}}^{(k, q-1)})_j\|_{H^0}\,,
\end{equation}

\noindent
where  $ \| (V_{I_{k,q}}^{(k, q-1)})_j\|_{H_0}$ will turn out to be bounded in the next step.
Regarding estimate (\ref{Snorm}), it follows from the following computation:
\begin{eqnarray}
& &\|(S_{I_{k,q}})_j\|\label{S-in}\\
&\leq& 2\left\|\frac{1}{G_{I_{k,q}}-E_{I_{k,q}}}P^{(+)}_{I_{k,q}}\,(V^{(k,q-1)}_{I_{k,q}})_j\,P^{(-)}_{I_{k,q}}\right\|\\
&=&2\left\|\frac{1}{G_{I_{k,q}}-E_{I_{k,q}}}P^{(+)}_{I_{k,q}}(H_{I_{k,q}}^0+1)^{\frac{1}{2}}(H_{I_{k,q}}^0+1)^{-\frac{1}{2}}(V^{(k,q-1)}_{I_{k,q}})_j(H_{I_{k,q}}^0+1)^{-\frac{1}{2}}P^{(-)}_{I_{k,q}}\right\| \\
&\leq&2\left\|\frac{1}{G_{I_{k,q}}-E_{I_{k,q}}}P^{(+)}_{I_{k,q}}(H_{I_{k,q}}^0+1)^{\frac{1}{2}}\right\|   \|(V_{I_{k,q}}^{(k,q-1)})_j\|_{H^0}\\
&\leq & \frac{2\sqrt{2}}{\Delta_{I_{k,q}}}\|(V_{I_{k,q}}^{(k,q-1)})_j\|_{H^0}\,, \label{S-fin}
\end{eqnarray}
where we have used (\ref{op-norm-G-2}) for the last inequality.

\noindent
Analogously, making use of (\ref{op-norm-G}) and $(H_{I_{k,q}}^0+1)^{\frac{1}{2}}P^{(-)}_{I_{k,q}}=P^{(-)}_{I_{k,q}}$, we estimate
\begin{equation}\label{S-norm-2}
\|(S_{I_{k,q}})_{j}(H^0_{I_{k,q}}+1)^{\frac{1}{2}}\|=\|(H^0_{I_{k,q}}+1)^{\frac{1}{2}}(S_{I_{k,q}})_{j}\|\leq \frac{2+\sqrt{2}}{\Delta_{I_{k,q}}} \|(V_{I_{k,q}}^{(k, q-1)})_{j}\|_{H^0}\,.
\end{equation}
Next, we want to prove that
\begin{eqnarray}
\|(V^{(k,q-1)}_{I_{k,q}})_j\|_{H^0}&\leq&\label{V-ineq}\\
\sum_{p=2}^{j}\,\frac{(2c)^p}{p!}&&\sum_{ r_1\geq 1 \dots, r_p\geq 1\, ; \, r_1+\dots+r_p=j}\,\|\,(V^{(k,q-1)}_{I_{k,q}})_{r_1}\|_{H^0}\|\,(V^{(k,q-1)}_{I_{k,q}})_{r_2}\|_{H^0}\dots \|\,(V^{(k,q-1)}_{I_{k,q}})_{r_p}\|_{H^0}\nonumber \\
+2\|V^{(k,q-1)}_{I_{k,q}}\|_{H^0}&& \sum_{p=1}^{j-1}\,\frac{(2c)^p}{p!}\,\sum_{ r_1\geq 1 \dots, r_p\geq 1\, ; \, r_1+\dots+r_p=j-1}\,\|\,(V^{(k,q-1)}_{I_{k,q}})_{r_1}\|_{H^0}\|\,(V^{(k,q-1)}_{I_{k,q}})_{r_2}\|_{H^0}\dots \|\,(V^{(k,q-1)}_{I_{k,q}})_{r_p}\|_{H^0}\,,\nonumber 
\end{eqnarray}
where $c:= \frac{2+\sqrt{2}}{\Delta_{I_{k,q}}}(> \frac{2\sqrt{2}}{\Delta_{I_{k,q}}})$.
In order to show this, we note that formula (\ref{formula-v_j-bis}) yielding $(V_{I_{k,q}}^{(k, q-1)})_j$ contains two sums. We first deal with the second, namely
$$\sum_{p\geq 1, r_1\geq 1 \dots, r_p\geq 1\, ; \, r_1+\dots+r_p=j-1}\frac{1}{p!}\text{ad}\,(S_{I_{k;q}})_{r_1}\Big(\text{ad}\,(S_{I_{k;q}})_{r_2}\dots (\text{ad}\,.(S_{I_{k,q}})_{r_p}(V^{(k,q-1)}_{I_{k,q}})\Big)\,. $$
Each summand of the above sum is in turn a sum of $2^p$ terms which, up to a sign,  are permutations of
$$(S_{I_{k,q}})_{r_1}(S_{I_{k,q}})_{r_2}\ldots (S_{I_{k,q}})_{r_p}V_{I_{k,q}}^{(k, q-1)}\,,$$
with the potential $V_{I_{k,q}}^{(k, q-1)}$ allowed to appear at any position.
It suffices to study only one of these terms, for the others can be treated in the same way. For instance, we can 
treat $$(S_{I_{k,q}})_{r_1} V_{I_{k,q}}^{(k, q-1)}   (S_{I_{k,q}})_{r_2}\ldots (S_{I_{k,q}})_{r_p}.$$
Notice that
\begin{eqnarray}
&&\|(S_{I_{k,q}})_{r_1} V_{I_{k,q}}^{(k, q-1)}   (S_{I_{k,q}})_{r_2}\ldots (S_{I_{k,q}})_{r_p}\|_{H^0} \nonumber \\
&=&\|(H_{I_{k,q}}^0+1)^{-\frac{1}{2}} (S_{I_{k,q}})_{r_1}  (H_{I_{k,q}}^0+1)^{\frac{1}{2}}   (H_{I_{k,q}}^0+1)^{-\frac{1}{2}} V_{I_{k,q}}^{(k, q-1)}(H_{I_{k,q}}^0+1)^{-\frac{1}{2}}(H_{I_{k,q}}^0+1)^{\frac{1}{2}}    (S_{I_{k,q}})_{r_2}\ldots (S_{I_{k,q}})_{r_p}  (H_{I_{k,q}}^0+1)^{-\frac{1}{2}}\|  \nonumber \\
&\leq& \|V_{I_{k,q}}^{(k, q-1)} \|_{H^0} \|(S_{I_{k,q}})_{r_1}(H_{I_{k,q}}^0+1)^{\frac{1}{2}} \|\, \|(H_{I_{k,q}}^0+1)^{\frac{1}{2}}    (S_{I_{k,q}})_{r_2}\| \ldots \|(S_{I_{k,q}})_{r_p}\| \nonumber\\
&\leq &c^p \|V^{(k,q-1)}_{I_{k,q}}\|_{H^0} \|\,(V^{(k,q-1)}_{I_{k,q}})_{r_1}\|_{H^0}\|\,(V^{(k,q-1)}_{I_{k,q}})_{r_2}\|_{H^0}\dots \|\,(V^{(k,q-1)}_{I_{k,q}})_{r_p}\|_{H^0}\,,\nonumber
\end{eqnarray}
where (\ref{Snorm}) and (\ref{S-norm-2}) have been used. 
\noindent
Putting these terms together we get the second sum of (\ref{V-ineq}).\\

As for the first sum in (\ref{formula-v_j-bis}), i.e.,

$$\sum_{p\geq 2, r_1\geq 1 \dots, r_p\geq 1\, ; \, r_1+\dots+r_p=j}\frac{1}{p!}\text{ad}\,(S_{I_{k,q}})_{r_1}\Big(\text{ad}\,(S_{I_{k,q}})_{r_2}\dots (\text{ad}\,(S_{I_{k,q}})_{r_p}(G_{I_{k,q}}) \Big)\,,$$
we note that each of its summands is in turn the sum up to a sign of all permutations of
$$(S_{I_{k,q}})_{r_1}(S_{I_{k,q}})_{r_2}\ldots(S_{I_{k,q}})_{r_{p-1}}[-P_{I_{k,q}}^+(V_{I_{k,q}}^{(k,q-1)})_{r_p} P_{I_{k,q}}^- - P_{I_{k,q}}^-(V_{I_{k,q}}^{(k,q-1)})_{r_p}P_{I_{k,q}}^+]\,.$$ 
Now a very minor variation of the computations above shows that the $\|\cdot\|_{H^0}$-norm of the first sum  in  (\ref{formula-v_j-bis}) is bounded
from above by 
$$\sum_{p=2}^{j}\,\frac{(2c)^p}{p!}\sum_{ r_1\geq 1 \dots, r_p\geq 1\, ; \, r_1+\dots+r_p=j}\,\|\,(V^{(k,q-1)}_{I_{k,q}})_{r_1}\|_{H^0}\|\,(V^{(k,q-1)}_{I_{k,q}})_{r_2}\|_{H^0}\dots \|\,(V^{(k,q-1)}_{I_{k,q}})_{r_p}\|_{H^0}\,;
$$
here we have implicitly assumed that $c>1$, without loss of generality.

From now on,  we closely follow the proof of Theorem 3.2 in \cite{DFFR}; that is, assuming $\|V^{(k,q-1)}_{I_{k,q}}\|_{H_0}\neq 0$, we recursively define numbers $B_j$, $j\geq 1$, by the equations
\begin{eqnarray}
B_1&:= &\|V^{(k,q-1)}_{I_{k,q}}\|_{H^0}=\|(V^{(k,q-1)}_{I_{k,q}})_1\|_{H^0} \label{B1} \,,\\
B_j&:=&\frac{1}{a}\sum_{k=1}^{j-1}B_{j-k}B_k\,,\quad j\geq 2\,, \label{def-Bj}
\end{eqnarray}
with \,$a>0$\, satisfying the relation
\begin{equation}\label{a-eq}
e^{2c a}-1+ \left( \frac{e^{2c a}-2c a -1}{a}\right)-1=0
\end{equation}
Using (\ref{B1}), (\ref{def-Bj}), (\ref{V-ineq}),  and an induction, it is not difficult to prove that (see Theorem 3.2 in \cite{DFFR}) for $j\geq 2$
\begin{equation}\label{bound-V-B}
\|(V^{(k,q-1)}_{I_{k,q}})_j\|_{H^0}\leq B_j\,\Big(\frac{e^{2c a}-2c a-1}{a}\Big)+2\|V^{(k,q-1)}_{I_{k,q}}\|_{H^0}\,B_{j-1}\Big(\frac{e^{2c a}-1}{a}\Big)\,.
\end{equation}
From (\ref{B1}) and (\ref{def-Bj}) it also follows that
\begin{equation}\label{bound-b}
  B_j\geq \frac{2B_{j-1}\|\,V^{(k,q-1)}_{I_{k,q}}\,\|_{H^0}}{a}\,\quad \Rightarrow\quad B_{j-1}\leq a\frac{B_j}{2\|\,V^{(k,q-1)}_{I_{k,q}}\,\|_{H^0}}\,,
\end{equation} 
which, when combined with (\ref{bound-V-B}) and (\ref{a-eq}), yields
\begin{equation}\label{bound-b-bis}
B_j\geq \|\,(V^{(k,q-1)}_{I_{k,q}})_j\|_{H^0}\,.
\end{equation} 
The numbers $B_j$ are the Taylor's coefficients of the function
\begin{equation}
f(x):=\frac{a}{2}\cdot \left(\,1-\sqrt{1- (\frac{4}{a}\cdot \|V^{(k,q-1)}_{I_{k,q}}\|_{H^0}) \,x }\,\right)\,,
\end{equation}
(see  \cite{DFFR}). We observe that
\begin{eqnarray}
\|(V^{(k,q-1)}_{I_{k,q}})^{diag}_j \|_{H^0}&=& \max\{\|P^{(+)}_{I_{k;q}}(V^{(k,q-1)}_{I_{k,q}})_j P^{(+)}_{I_{k;q}}\|_{H^0}\,,\,\|P^{(-)}_{I_{k,q}}(V^{(k,q-1)}_{I_{k,q}})_j P^{(-)}_{I_{k,q}}\|_{H^0}\}\label{diag-est-in}\\
& =&\max_{\#=\pm }\,\|(\frac{1}{H^0_{I_{k,q}}+1})^{\frac{1}{2}}P^{(\#)}_{I_{k,q}}(V^{(k,q-1)}_{I_{k,q}})_j P^{(\#)}_{I_{k,q}}(\frac{1}{H^0_{I_{k,q}}+1})^{\frac{1}{2}}\|\\
&=&\max_{\#=\pm }\,\|P^{(\#)}_{I_{k,q}}(\frac{1}{H^0_{I_{k,q}}+1})^{\frac{1}{2}}(V^{(k,q-1)}_{I_{k,q}})_j (\frac{1}{H^0_{I_{k,q}}+1})^{\frac{1}{2}}P^{(\#)}_{I_{k,q}}\|\\
&\leq & \|(V^{(k,q-1)}_{I_{k,q}})_j \|_{H^0}\,.\label{diag-est-fin}
\end{eqnarray}
Therefore the radius of analyticity, $t_0$,  of 
\begin{equation}
\sum_{j=1}^{\infty}t^{j-1}\|(V^{(k,q-1)}_{I_{k,q}})^{diag}_j \|_{H^0}=\frac{d}{dt}\,\Big(\sum_{j=1}^{\infty}\frac{t^{j}}{j}\|(V^{(k,q-1)}_{I_{k,q}})^{diag}_j \|_{H^0}\Big)
\end{equation}
is bounded below by the radius of analyticity of $\sum_{j=1}^{\infty}x^jB_j$, i.e.,
\begin{equation}\label{radius}
t_0\geq \frac{a}{4\|V^{(k,q-1)}_{I_{k,q}}\|_{H^0}}\geq \frac{a}{4}
\end{equation}
where we have assumed $0<t<1$ and invoked  the assumption $\|V^{(k,q-1)}_{I_{r,i}}\|_{H^0}\leq t^{\frac{r-1}{4}}$.
Thanks to the inequality in (\ref{Snorm}), the same bound holds true for the radius of convergence of the series $S_{I_{k,q}}:=\sum_{j=1}^{\infty}t^j(S_{I_{k,q}})_j\,$\,.
For $0<t<1$ and in the interval $(0,\frac{a}{8})$, by using (\ref{B1}) and  (\ref{bound-b-bis}) we can estimate
\begin{eqnarray}
\sum_{j=1}^{\infty}t^{j-1}\|(V^{(k,q-1)}_{I_{k,q}})^{diag}_j \|_{H^0}&\leq &\frac{1}{t}\sum_{j=1}^{\infty}t^jB_j\\
&=&\frac{1}{t}\cdot \frac{a}{2}\cdot \left(\,1-\sqrt{1- (\frac{4}{a}\cdot \|V^{(k,q-1)}_{I_{k,q}}\|_{H^0}) \,t }\,\right)\\
&\leq &(1+C_a \cdot t )\,\|V^{(k,q-1)}_{I_{k,q}}\|_{H^0}
\end{eqnarray}
for some $a$-dependent constant $C_a>0$.
Hence the inequality in (\ref{bound-V})  holds true, provided  $t$ is sufficiently small, independently of $N$, $k$, and $q$. 
In a similar way, we derive (\ref{bound-S}) and (\ref{S-Hest}), using  (\ref{S-in})-(\ref{S-fin}) and (\ref{S-norm-2}), respectively. \qed

\end{appendix}

\end{document}